\documentclass[twocolumn]{revtex4-1}

\usepackage{graphicx}

\usepackage{tabularx,ragged2e,booktabs,caption}

\graphicspath{{./figs/}} % Path to graphics directory

\usepackage{tikz}
\usepackage{bm}

\definecolor{mycoljd}{rgb}{0.3686, 0.3098, 0.6353}
\definecolor{mycolgot}{rgb}{0.2941 0.5447 0.7494}

\definecolor{mycolkidas}{rgb}{0.6859 0.4035 0.2412}

\definecolor{mycolwangprl2012}{rgb}{1.000 0.5482 0.1}

\definecolor{mycolwangjt}{rgb}{0.3718 0.7176 0.3612}

\definecolor{mycolwangpre}{rgb}{0.8650 0.8110 0.4330}

\definecolor{mycolwangprl2013}{rgb}{0.9942 0.7583 0.4312}

\definecolor{mycolwanghst2018}{rgb}{0.9047 0.1918 0.1988}

\definecolor{mycolsar1}{rgb}{0.64 0.64 0.64}

\definecolor{mycolwangtherm}{rgb}{0.6365 0.3753 0.6753}

\definecolor{mycolkidad}{rgb}{0.9718 0.5553 0.7741}

\definecolor{mycolsar2}{rgb}{0.5 1.0 0.0}
\def\fs{\footnotesize}

\renewcommand{\vec}[1]{\boldsymbol{#1}}
\def\re {R_\lambda}
\def\reL {R_L}
\def\mt {M_t}
\def\mtd {M_{d}}

\def\avdis {\langle\epsilon\rangle}
\def\avdiss {\langle\epsilon_s\rangle}
\def\avdisd {\langle\epsilon_d\rangle}

\def\be {\begin{equation}}
\def\ee {\end{equation}}

\newcommand{\req}[1]{Eq.\ \ref{eq:#1}}
\newcommand{\rfig}[1]{Fig.\ \ref{fig:#1}}
\newcommand{\rtab}[1]{Table~\ref{tab:#1}}

\newcommand{\colr}[1]{\textcolor{red}{{#1}}}

\def\ut {\vec{u}}
\def\us {\vec{u}_s}
\def\ush {\vec{\hat{u}}_s}
\def\ud {\vec{u}_d}
\def\ps {p_s}
\def\pd {p_d}
\def\urms {u_\textrm{rms}}
\def\prms {p_\textrm{rms}}
\def\rhorms {\rho_\textrm{rms}}
\def\Trms {T_\textrm{rms}}
\def\rhom {\la\rho\ra}
\def\rhorms {\rho_\textrm{rms}}
\def\usrms {u_{s,\textrm{rms}}}
\def\udrms {u_{d,\textrm{rms}}}
\def\psrms {p_{s,\textrm{rms}}}
\def\pdrms {p_{d,\textrm{rms}}}
\def\pm {\la p \ra}
\def\Tm {\la T \ra}

\def\DDE  {DDE}

\def\la {\langle}
\def\ra {\rangle}

\def\D {{\cal D}}

\def\deltai {\delta_\infty}

\begin{document}

\title{Universality and scaling in compressible turbulence}

%\author{
%Diego A.\  Donzis\affil{1}
%{Texas A\&M University, College Station, USA} 
%John Panickacheril John\affil{1}
%{Texas A\&M University, College Station, USA} 
%}
\author{Diego A. Donzis}
\email[]{donzis@tamu.edu}
\affiliation{Department of Aerospace Engineering, Texas A\&M University, College Station, Texas 77843, USA}
\author{John Panickacheril John}
%\email[]{johnthampu@tamu.edu}
\affiliation{Department of Aerospace Engineering, Texas A\&M University, College Station, Texas 77843, USA}

%\contributor{
%Submitted to Proceedings of the National Academy of Sciences
%of the United States of America
%}

%%%Newly updated.
%%% If significance statement need, then can use the below command otherwise just delete it.
%\significancetext{
%Complex non-linear dissipative systems with a wide range of 
%spatial and temporal interacting scales are notoriously 
%difficult to characterize and understand. A prime example is 
%turbulence which, when coupled with thermodynamic fluctuations, 
%is critical in countless natural and engineering systems.
%Our work provides a different perspective on how
%scaling laws and thus models  can be constructed,
%}

%\begin{article}
\begin{abstract}
Turbulent flows, ubiquitous in nature and engineering,
comprise fluctuations over a
wide range of spatial and temporal scales.
While flows with fluctuations in thermodynamic variables
are much more common, much less is known about these 
flows than their incompressible counterparts in which 
thermodynamics is decoupled from hydrodynamics.
A critical element in the study of the latter 
has been the concept of universal scaling 
laws which 
%resulted in critical advances, 
%not only for the study of turbulence but also for other
%fields.
provides fundamental as well as practical information 
about the spatio-temporal behavior of these complex systems.
Part of this success is due to the dependence on
a single non-dimensional parameter, that is the 
Reynolds number.
Universality in 
compressible flows, on the other hand, 
%require 
%more parameters to characterize its statistical 
%behavior.
%Universality has thus been 
have proven to be 
more elusive as no unifying set of 
parameters were found to yield universal scaling laws.
This severely limits our understanding of these flows 
and the successful development of theoretically 
sound models.
Using results in specific asymptotic limits 
of the governing equations
we show that universal scaling is indeed observed 
when the set of governing 
parameters is expanded to include
internally generated dilatational scales
which are the result of the driving mechanisms that produce the 
turbulence.
The analysis demonstrate why previous scaling laws fail in the
general case, and opens up new venues 
to identify physical processes of
interest and create turbulence models needed for 
simulations of turbulent flows at realistic conditions.
We support our results 
with a new massive database of highly-resolved 
direct numerical simulations 
along with all data available in the literature.
In search of universal features, we suggest classes
which bundle the evolution 
of flows in the new parameter space.
An ultimate asymptotic regime
predicted independently by renormalization 
group theories and statistical mechanics is also 
assssed with available data.
\end{abstract}

\maketitle

%\keywords{turbulence | compressibility | universality | scaling}

%\abbreviations{SAM, self-assembled monolayer; OTS, octadecyltrichlorosilane}

%\dropcap{M}
Most flows in nature and engineering are turbulent
exhibiting fluctuations in velocity, pressure and virtually
all hydrodynamic and thermodynamic variables 
describing the fluid state.
A strong simplification widely used is the assumption of 
incompressibility, that is, the volume preservation of 
fluid elements as they deform and rotate in space and time.
This condition which implies a solenoidal velocity field
and constant density fields,
also commonly leads to a decoupling of hydrodynamics and 
thermodynamics, simplifying the problem greatly.
However,
this is not the case in the most general condition which
includes flows in astrophysical and cosmological contexts, 
aerodynamics of aircraft and spacecraft, geophysical 
phenomena such as volcanic eruptions, and reacting 
turbulent flows, among many others. 
Due to the extreme complexity of these systems
with their wide range of non-linearly interacting 
spatial and temporal scales, 
it is only in a statistical sense that these flows
are expected to be characterized and understood.
Even then, a main obstacle in this endeavor 
is the apparent difficulty seen in the literature
to find universal scaling laws for 
different flows or for similar flows
under diverse conditions. 
These attempts are commonly based on 
two non-dimensional parameters, namely, the
Taylor Reynolds number ($\re$) 
which represents the relative 
importance of inertial to viscous effects, and the 
turbulent Mach number 
($\mt$) 
which is the ratio
of a characteristic velocity to the speed of sound. 
However, these efforts have remained generally 
inconclusive and 
support for different proposals has been limited
and often contradictory. 
In most cases, scaling laws are hard to discern 
and found to depend on the 
type of driving mechanism that sustains turbulence 
\cite{KGFK2012,FKS2008,FKS2009,WJWMCSXCCS2018,CGTFJFM2019,WYSXHC2013,WGWPRF2017},
initial conditions \cite{PG2004,SPK2001} 
or other details of the flow.
This is in contrast to the relative success %of universal scaling
in incompressible turbulence where %at least to second-order,
universal scaling laws have proven insightful and practically 
useful \cite{frisch95,sreeni2018}.
% Scaling and universality have also been tremendously 
% useful 
% % in a wide range of natural 
% %and man-made systems such as 
% in rock fracture \cite{DSD2007}, economics \cite{SP2001},
% medicine \cite{IRPM+1998}, earthquakes \cite{corral2004},
% %urban spaces \cite{CP2004}, 
% cloning \cite{RLC+2018}, 
% among many others.

As we argue here, universal scaling in compressible 
turbulence is indeed observed, but only when internally 
generated scales representing purely
compressible motions are the basis 
for similarity scaling.
This represents a departure from previous efforts
which focus on externally imposed scales
leading to an incomplete set 
of governing non-dimensional parameters to 
define the state of turbulence.
%and, thus, a lack of guidance as to what type of
%statistical equilibrium the flow is expected to achieve 
%at different flow conditions.
We show how this state of affairs followed
from a few (sometimes 
implicit) assumptions about the variables that control 
the system.

In particular we show that
to define the state of turbulence,
dilatational motions 
cannot be neglected, as traditionally done.
%\colr{[JPJ: I was thinking if we 
%can rephrase the sentence before a bit, people dont say diltattional 
%motions are not important but they havenet figured out a way to include 
%dilational motions in a general context for all flows for a wider range of circumstance as we do here. Usually their studies were 
%so specfic that $\delta$ was similar in their particular study so ofcourse they negelect $\delta$ always and be consistent in that 
%particular study but ofcourse inconsistent with their other studies for which they say dilataional motions might be the 
%difference but have never attempted to quantify it or explain in a 
%general context as we do here. Just a thought]}  
Using first pressure fluctuations, we show that the flow 
undergoes a transition between two equilibrium states,
one dominated by incompressible non-linear dynamics 
and the other one by compressible or dilatational motions
with simple linear dynamics.
The scaling on these two limits depends on 
different set of parameters, one including dilatational
motions while the other not.
The relative importance of these two equilibria
lead naturally to governing parameters
that uniquely define the statistical 
state of fluctuating thermodynamic variables.
Classical parameters are also shown to fail 
to collapse other quantities important to the dynamics 
of turbulence such the rate at which turbulent 
kinetic energy 
is dissipated as well as the rate at which vortical motions
are produced which exemplify both the non-linear behavior 
and conspicuous non-Gaussianity of turbulence.
We show that universality is also 
unraveled for these variables 
by the introduction of dilatational motions into 
the governing parameters.
These, in turn, define statistical regimes in the extended
parameter space which help explain discrepancies
in the literature and explain when and in what sense 
universal scaling is expected.
The results presented, thus, provide a 
new perspective for compressible turbulence, 
and perhaps more general 
non-linear multiscale systems.

\section{Background}

A common paradigm to understand 
fundamental issues in turbulence is to focus on 
the intrinsic dynamics of
statistically steady isotropic,
homogeneous flows to avoid complications from 
wall effects, spatial non-uniformities and transients.
Even in this simplified setup,
the extraordinary difficulties associated with 
rigorous treatments of the non-linear 
governing equations (that is, the Navier-Stokes
equations)
have resulted in 
experiments and numerical simulations being 
a main driver of progress.
In the case of compressible turbulence,
experiments under these simplifying assumptions 
in well controlled conditions 
are also exceedingly difficult and thus scarce.
Therefore, 
numerical simulations have become a main tool in investigating
these flows.
Of particular importance %in fundamental studies of turbulence 
are the so-called 
direct numerical simulations (DNS) where the exact governing 
equations expressing conservation of mass, momentum and energy
are solved such that all temporal and spatial scales 
are accurately resolved by the numerical scheme \cite{MM98,IGK2009}.
Due to 
the complexity of these equations and the wide range of 
scales, DNS are computationally extremely 
expensive and require massive computational power.
This is the type of simulations we will use here which, 
to the authors' knowledge, is among 
the largest and best-resolved databases of
isotropic compressible 
turbulence in the published literature.

Because of the comparatively larger body of  literature 
in incompressible turbulence and its more limited
set of governing parameters and physical processes,
a common approach to study 
compressible flows is to unveil differences from 
incompressible counterparts by quantifying
departures from known scaling laws.
This has led to 
significant advances in the field 
\cite{lele1994,GB2009}.
A practical, yet rigorous, approach
to separate aspects associated with incompressible 
and compressible motions is the 
Helmholtz decomposition in which the velocity field
is split as $\vec{u}=\us + \ud$ where 
$\us=\nabla\times \vec{A}$ ($\vec{A}$ is the vector potential) and
$\ud=\nabla\phi$ ($\phi$ is the scalar potential). 
It is trivial to show 
that $\vec{u}_s$ is solenoidal (i.e.\ $\nabla\cdot \us=0$)
and that $\ud$ is irrotational (i.e.\ $\nabla\times \ud=0$).
Since $\ud$ is identically zero in incompressible flows,
its existence is ascribed to purely compressibility effects.
It is important to note, though, that compressibility can also 
affect $\us$. While the decomposition of 
related variables is also possible 
(e.g.\ $\sqrt{\rho}\vec{u}$ \cite{KO1990})
we have verified that conclusions here are independent 
of this choice.

Decades of accumulated numerical work,
have made some aspects of these flows 
increasingly clear.
A few examples include 
mixing inhibition due to compressibility 
\cite{sarkarjfm1995,vsljfm1996,ni2016},
changes in decay rates due to initial 
level of dilatation or thermodynamic fluctuations
\cite{SPK2001,praturi2019effect,BMR1993,RB1997,SEHK1991},
and,
when one compares across studies as done here,
the failure of traditional parameters 
(in particular Reynolds and turbulent Mach numbers)
to provide universal descriptions these flows.
%\cite{FKS2008,FKS2009,SFH+2009,SFK2008}.
%This is particularly evident when one compares
%different flows at the same 
%$\mt$,
%{=$u_{rms}/\langle c \rangle 
%\left(\textrm{ where } u_{rms} = u_{i} u_{i}\right) $}
%as done below.

The purpose of this work is 
to assess in which sense compressible turbulence
can be dubbed universal and under what 
conditions one can expect universal scaling.
In doing so we propose a set of parameters that 
define new regimes and lead 
to universal scaling laws that 
can, indeed, collapse all the data available in the literature.
This leads to both the recognition of limitations 
of current approaches as well as new paths for more
general understanding and modeling approaches.

\section{Governing parameters, scaling and similarity}

In order to explore similarity scaling in compressible
flows, we first discuss key fundamental issues 
involved in its application to  
their incompressible counterparts. 
In this case the equations for conservation of mass and momentum,
known as the Navier-Stokes equations,
are sufficient to describe the phenomenon to great accuracy.
In search of similarity, one can use a characteristic
velocity ($\cal U$) and length scale ($\cal L$)
to normalize all variables 
in the governing equations. This results in a single similarity
parameter, the Reynolds number $R={\cal U} {\cal L}/\nu$
where $\nu = \mu/\rho$ with $\mu$ being the dynamic viscosity
and $\rho$ the (constant) density. 
Geometrically similar flows are then expected to have 
identical properties if the Reynolds number is the same. 
In principle, 
if scaling laws in $R$ are known 
and $\cal U$ and $\cal L$ are known parameters of the 
problem at hand
(e.g.\ mean speed and the mesh size in grid turbulence, or 
the size of an object generating a turbulent wake) 
then one obtains 
useful predictive capabilities by simply knowing 
the geometry and conditions of the flow.
Since ${\cal L}$ and ${\cal U}$ are externally imposed 
we can call this, {\it external} similarity.
%The expectation of external similarity
%finds support in classical phenomenology of 
%turbulence in a specific flow configuration. 
%In it, one can postulate the existence of 
%small-scale universality
%based on the so called cascade of energy:
%energy is produced or introduced
%at large scales and through a 
%successions of instabilities it is transferred 
%to increasingly small scales where ultimately energy 
%is dissipated by viscous effects.
%This results in a universal behavior at scales 
%much smaller than the largest scales 
%in the flow. 
%Thus, it is some characteristics 
%of the large scales that determine to some degree
%the behavior of the wide range of scales that develop.
%In other words, 
%small-scale statistics depend only on the large-scale 
%features of the flow, characterized 
%by $\cal L$, $\cal U$, and the only parameter in the 
%equations of motions, namely, the viscosity of the fluid $\nu$. 
Dimensional analysis would then imply that 
an appropriately normalized statistics 
of interest $Q$, can be represented as
$\overline{Q}=f_i(R)$
where an overbar denotes normalization and
$f_i$ is presumably a universal function
for incompressible flows.

This approach, however, proves to be limiting when
comparisons {\it across different flows} is attempted.
For example, the normalized mean dissipation rate 
$\avdis {\cal L}/{\cal U}^3$ ($\epsilon$ is the instantaneous
dissipation rate per unit mass and angular brackets are
suitably defined ensemble
averages) is not 
expected to be the same for a body wake or grid turbulence 
if ${\cal L}$ is the diameter of the body 
for the former and the grid spacing for the latter
though the order of magnitude may be correctly predicted. 
To compare across flow, thus, it is common to 
use the large-scale Reynolds number 
$\reL=\urms L/\sqrt{3}\nu $, 
where $\urms^2=\la|\vec{u}-\la \vec{u}\ra|^2\ra$ 
(angular brackets are
suitably defined ensemble averages) 
is the root-mean-square velocity 
and $L$ the integral length scale, or the Taylor Reynolds 
number $\re=\urms\lambda/\sqrt{3}\nu$ where $\lambda$ is the Taylor
microscale which is an intermediate scale between the largest scales 
and the smallest dissipative scales. 
A vast literature on incompressible turbulence 
has been devoted to elucidating
the scaling of different statistical features 
of turbulence with $\re$, that is:
\be
\overline{Q}=f_i(\re)
\label{eq:iss}
\ee
%For multiple-point statistics, such as the so-called
%structure functions, one generalizes the above 
%expression to, for example, $\overline{Q}=f(\re,r/L)$ for a 
%two-point statistic with the two locations 
%separated by a distance $r$. In a number of instances,
%\req{iss} is expected to be a case of similarity 
%of the first kind \cite{BZ1972} for high $\re$, namely 
%$\lim_{\re\rightarrow\infty}f(\re) = C$ where $C$ is a
%(potentially universal) constant. 
%This is the case of well-studied quantities such as 
%the normalized dissipation rate \cite{sreeni1984} or
%the skewness of velocity gradients which is a measure of
%vorticity production due to non-linear mechanisms 
%\cite{SA97}.

Two critical issues are worth noting here.
First, there is the implicit assumption that a single
velocity scale and a single length scale are 
enough to completely characterize the flow, at least,
in a statistical sense. 
%This, in general, is appropriate when the Navier-Stokes
%equations are forced in a narrow compact 
%support in Fourier space.
Second, these flow scales 
($\urms$, $L$ and $\lambda$) are
computed from the flow itself and their value can only 
be estimated to within order of magnitude 
from a priori known geometrical details of the flow setup.
One is thus forced to acknowledge the knowledge gap between 
a priori characteristic velocities and lengths from the original 
geometry %($\cal U$, $\cal L$)
and the resulting {\it internal} scales %that result 
from the flow dynamics subjected to those particular initial or
boundary conditions defined by ${\cal U}$ and ${\cal L}$.
When using internally generated characteristic scales,
we call this {\it internal} or {\it self} similarity.
While a disadvantage of such an approach is the lack of
pure predictive capabilities from a priori known characteristics
of the flow, its advantage lies on its ability 
of unravel universal aspects across geometrically 
different flows.

A well-known example is the phenomenology of 
Kolmogorov \cite{K41} who suggested internally 
generated scales (the Kolmogorov scales) 
%characteristic of small-scale motions
as a way to find universality. 
%(Under some 
%further assumptions this internal scales can be %readily
%estimated from $\urms$ and $L$.)
This universality
is indeed observed by the good
collapse 
of multi-point statistics 
(at least for low orders \cite{frisch95}) 
when normalized by Kolmogorov scales \cite{MY.II}.
One of the most salient examples 
is the observed collapse of the normalized 
energy spectrum from a variety of
flows at different Reynolds numbers
\cite{SA97}.

Most efforts to find universality in compressible turbulence
have attempted  to use known 
results from incompressible flows and study departures and
differences as compressibility levels increase 
\cite{lele1994}. 
The degree of compressibilty is
commonly measured by the Mach number constructed 
by a characteristic velocity $\cal U$, and the speed 
of acoustic propagation $c$, that is ${\cal M}\equiv  {\cal U} / c$. 
Clearly for incompressible flows, ${\cal M}=0$ since $c\rightarrow \infty$.
A well-known example of this paradigm 
is the widely studied reduction of
the spreading rate of a mixing layer with $\cal M$ 
(defined with some average convective velocity) 
\cite{SD.book.2006}.
%relative to the incompressible counterpart. 
There is still debate about
what the correct definition of $\cal M$ is, though 
\colr{\cite{hall1993experiments,dimo1991turb,freund2000,lele1994,mm}}.
%\textcolor{green}{added few references, should be enough}
Given the complexity of turbulence in general and compressible 
turbulence in particular,
a large body of literature has been devoted instead to
homogeneous isotropic flows, where no mean velocity exists.
This approach avoids the additional complexity of geomertical 
factors and allows for fundamental understanding of
intrinsic characteristics that emerge from the governing 
equations.
In such flows, which are the focus of this work as well,
it is common to characterize the degree of compressibility 
with the turbulent Mach number $\mt\equiv \urms/c$ where 
$c$ is the mean speed of sound. This clearly corresponds
to self similarity as described above.
An implicit assumption here is that the addition of 
the propagation speed of acoustic waves
provides a complete set of governing 
parameters of the flow. 
%Since $c$ is identified to the speed of sound only 
%when the dynamics is linearized, then  
Perhaps a more appropriate interpretation 
%of $c$ 
is apparent from the relation
$c^2\sim \la T\ra \sim \la p\ra/\la \rho\ra$ 
for a perfect gas.
Thus, $c$ completes the specification of the 
thermodynamic state of the 
flow (at least in a mean sense) when density is included 
in the set of governing parameters.
Note also that using the {\it mean} speed of sound amounts to
again seeking internal similarity as $c$, in general, depends on 
the flow solution which involves temperature 
fluctuations.
Self-similar scaling would then imply that there are
universal functions
$f_c$ for different non-dimensional quantities $\overline{Q}$
in the form%
\footnote{A more general list of parameters
would include
the ratio of specific heats, $\gamma$ but is not
included here for simplicity.}
\be
\overline{Q}=f_c(\re,\mt) .
\label{eq:css}
\ee
This has been a basic assumption in substantial 
amount of research of compressible flows.
For example, 
several theories have been proposed to determine the
scaling of so-called 
dilatational dissipation \cite{GB2009}
or the spectrum \cite{JD2016,WGWPRF2017,WJWMCSXCCS2018,xxx} with $\re$ and
$\mt$.
Weaker formulations of the form 
$\overline{Q}=f_i(\re)f_c(\mt)$ are also common.
All these different proposals have been tested against 
numerical simulations with mixed success 
and no universal behavior has emerged. 
%have also been used to attempt
%to test different predictions though, as we show below,
%data do not agree with each other. 
Part of this state of affairs, we 
argue, is because of an incorrect identification 
of relevant non-dimensional groups to determine the statistical 
state of turbulence. 
While $\mt$ compares turbulence velocities
to the propagation speed of linear waves, 
it does not contain 
any information on, for example, the amount on energy 
in those compressible modes
that may 
propagate in a wave-like fashion.

%While our focus here is on
%statistically steady, homogeneous and
%isotropic flows for which internal similarity 
%expressed by \req{css} has been extensively sought,
%we will also assess predictions against 
%more complex scenarios, 
%such as flows with mean shear.

\section{The role of dilatational motions}
The inadequacy of \req{css} 
is evident by the qualitatively distinct behavior observed 
for turbulence 
when the nature of the forcing or initial conditions 
are changed. 
The structure and dynamics at both 
terrestrial and astrophysical conditions were found to
depend strongly on whether 
driving forces \cite[(check this refs)]{KO1990,SFH+2009,FKS2009}
or initial state of the flow \cite[(check this refs)]{SPK2001,PG2004} 
contain or not dilatational contributions.

%From a more  theoretical perspective, 
%consider 
Consider
mass conservation
\be
\frac{1}{\rho}\frac{D\rho}{Dt} = -\nabla\cdot\vec{u} \ ,
\label{eq:continuity}
\ee
with $D/Dt$ being the standard substantial derivative.
Using Helmholtz decomposition one can easily 
see that changes in density following a fluid element 
can only be due to 
%the dilatational component of
%the advecting velocity, 
$\ud$.
Thus, 
it is clear that one needs a dilatational velocity scale,
${\cal U}_d$, to properly normalize \req{continuity}.
Furthermore, in many situations, 
density and pressure are related through an isentropic
or polytropic relation \cite{DJ2013,JD2016}
%at least for conditions considered here, 
which would then 
imply pressure would also be 
governed by dilatational motions.

In fact, 
starting with the full Navier-Stokes equations, 
one can derive evolution equations for the individual components
$\us$ and $\ud$ \cite{EHKS1990} 
where the solenoidal component of velocity and pressure
satisfy the incompressible Navier-Stokes equation.
The equations for the compressible part of the velocity field
is obtained by subtracting the incompressible system 
from the full set of equations.
%and the dilatational components are the reminder, 
%that is $\ud=\vec{u}-\us$ and $p_d=p-\ps$. 
While these equations are coupled, the
time 
scales and length scales associated with the evolution 
of each component are expected to be different 
\cite{SEHK1991,ristorcelli1997}.
In fact for small fluctuations and low Mach numbers, 
it has been argued that the
dilatational component of velocity 
decouples from its solenoidal component.
More formally, and under a slightly different approach,
one can linearize the governing equations, assume
isentropic fluctuations, and
project the velocity field onto the Craya basis in 
Fourier space \cite{SCC1997}. In this reference frame, 
the third axis is defined along 
the wavenumber vector $\vec{k}$, and thus the dilatational velocity 
projects only along that direction. The solenoidal
component is divergence free which in Fourier space implies 
$\ush\cdot\vec{k}=0$ (a caret indicates Fourier tranformed variables),
and thus lies on the plane 
perpendicular to it. Then, to leading order, 
the dilatational velocity and pressure
evolve according to:
\begin{eqnarray}
\partial u_d/\partial t & = & c_0 \  k\  P_d  \nonumber \\
\partial P_d/\partial t & = & - c_0\ k\ u_d 
\label{eq:linear}
\end{eqnarray}
where $u_d$ can be written as a scalar since the direction is 
always along the wavenumber vector. The 
speed of sound (constant in this simplified case) and the wavenumber
are $c_0$ and $k$ respectively.
$P_d=\iota\pd/\rho_0c_0$ is a normalized pressure
with $\iota=\sqrt{-1}$.
The problem is now simple enough to accept 
an analytical solution which we will explore momentarily.
For now, it suffices to say that, at least to first 
order, the dilatational motions of the 
governing equations decouple from solenoidal motions
and their dynamics are determined by an interplay 
between dilatational velocity and pressure.
In this approximation, there is no clear 
reason to believe 
that a characteristic solenoidal velocity scale 
would be an appropriate scale for \req{linear}.
Similar conclusions can be drawn 
for the evolution of the different
modes based on the Kovasznay
decomposition \cite{SD.book.2006}.

\begin{figure} %[h]
\begin{center}
\includegraphics[clip,trim=30 45 38 35,width=0.44\textwidth]{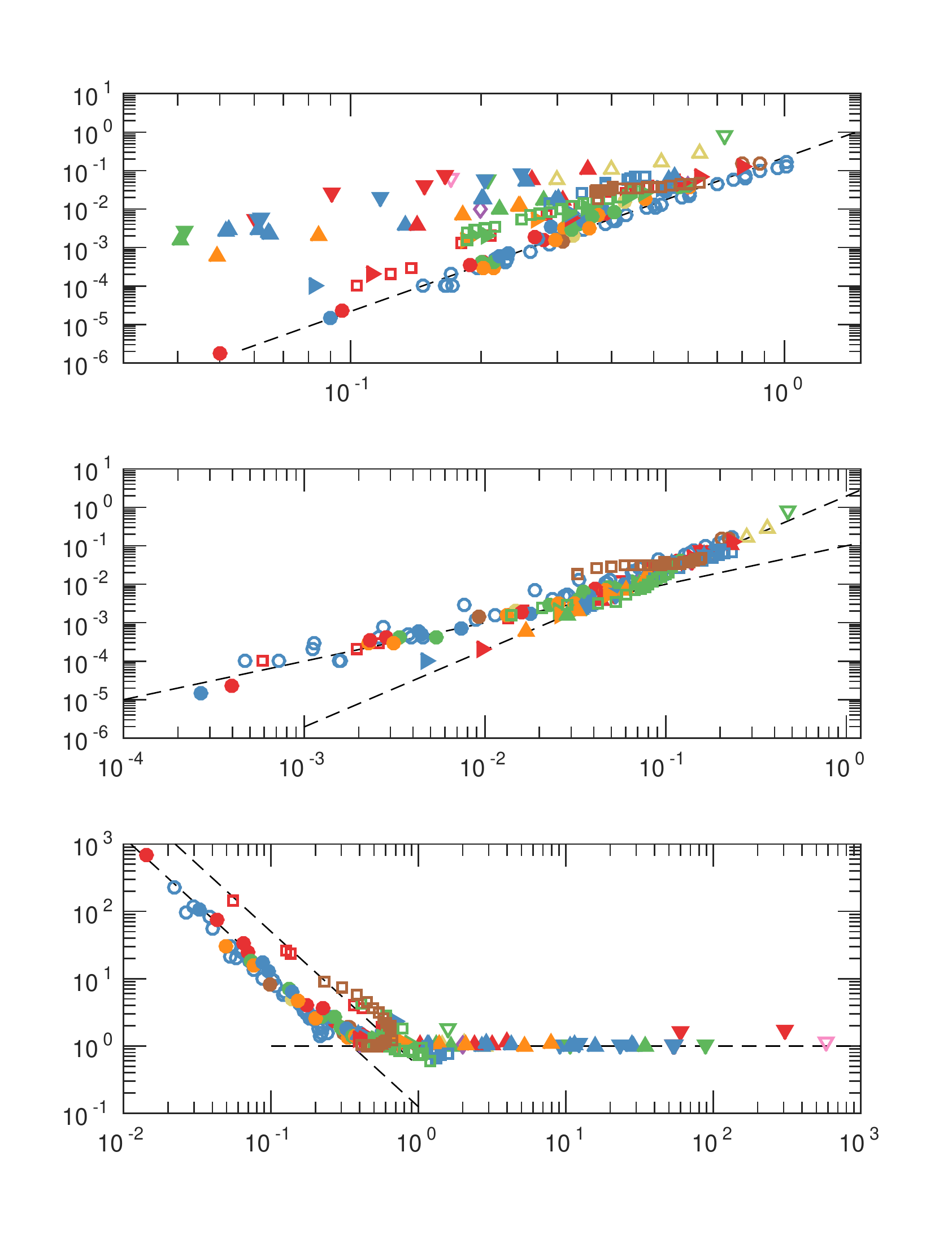}
\begin{picture}(0,0)
\put(-240,230){\rotatebox{90}{\normalsize{$\prms^2/\left(\la p\ra^2\right)$}}}
\put(-120,200){{$\mt$}}
\put(-240,140){\rotatebox{90}{\normalsize{$\prms^2/\la p\ra^2$}}}
\put(-120,95){{$\mtd$}}
\put(-240,2){\rotatebox{90}{\normalsize{$(\prms^2/\la p\ra^2) (F/
					\mtd^2 \gamma^2)$}}}
\put(-110,-3){$\cal D$}
\put(-175,75){\footnotesize{\rotatebox{-45}{$A=1.9$}}}
\put(-195,60){\footnotesize{\rotatebox{-45}{$A=1.0$}}}
\put(-44,24){\footnotesize{$F=1.0$}}
\end{picture}
\end{center}
\caption{
\label{fig:p2}
Scaling of variance of pressure with 
$\mt$ (a),  $\mtd$ (b) and $\cal D$ (c). 
 In (a), dashed lines are $\gamma^2 \mt^4/9$ \cite{DJ2013}.
In (b) dashed lines are $\gamma^2 \mtd^2/F_T$ (high $\mtd$) and 
$c \mtd$ for reference (low $\mtd$.) 
In (c) dashed lines are ${\cal D}^{-2}$ (low $\cal D$)
 for different values of $A$ and the asymptotic
 \DDE\ horizontal lines (high $\cal D$).
%\colr{[DD: why is one dashed line cyan?]}
%A: Current simulations(HIT solenoidal forced),
%B: Current simulations(HIT dilatational forced),
%C: Wang \textit{et al.}\cite{WYSXHC2013}(HIT dilatational forced)
%D: Wang \textit{et al.}\cite{WGWPRF2017}(HIT solenoidal forced)
%E: Sarkar \textit{et al.}\cite{SEHK1991}(HST anisotropic)
%F: Kida and Orszag\cite{KO1990,KO1992} (dilataional forced isotropic)
%G: Kida and Orszag\cite{KO1990,KO1992} (solenoidal forced isotropic)
%H: Sarkar\textit{et al.}\cite{SEH1992} (HST anisotropic)
%I: Wang\textit{et al.}\cite{WJWMCSXCCS2018} (dilataional forced isotropic) 
%J: Chen \textit{et al}\cite{CWLWC2018} (HST anisotropic)
}
\end{figure}
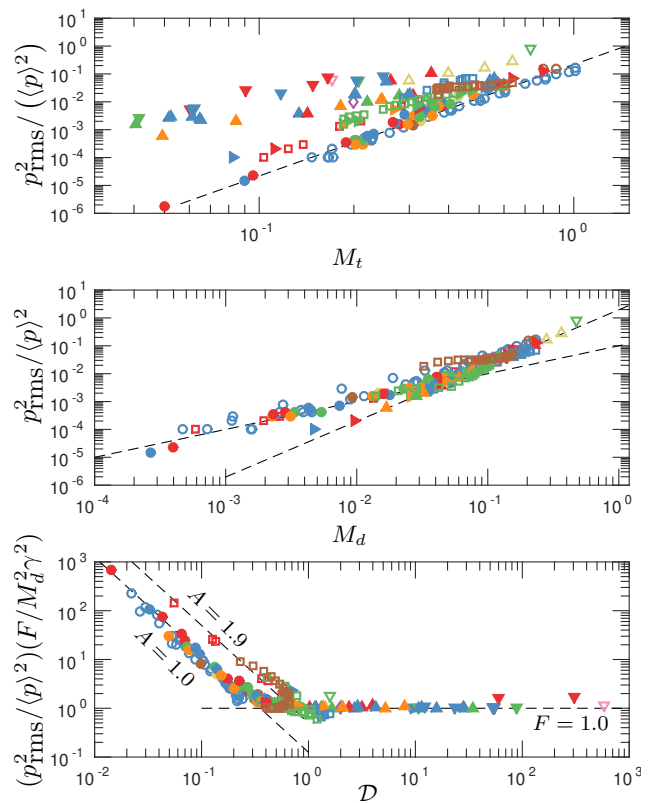

%This is consistent with the 
%results from studies available in the literature which
%highlight the importance of dilatational 
%motions and the necessity of expanding the parameter space to
%describe these flows.
%Unless dilatational
%motions imparted by the geometry or driving forces in the flow 
%are negligible,
%scaling relations in terms of $\re$ and $\mt$ alone are severely limited 
%in seeking universal relations. 
The accumulated data in the literature, in fact,
support these arguments which implies that 
a relation like \req{css} cannot hold in the general case.
An example is seen in \rfig{p2}(a) where we show the variance of 
pressure normalized by its mean in a statistically steady 
flow forced stochastically by purely solenoidal 
forcing (closed circles) and combination of solenoidal and dilatational
forcing (closed triangles).
Clearly dilatational forcing has a first order effect 
on the dynamics of the flow generating pressure fluctuations
orders of magitude larger than those seen at similar $\mt$
with solenoidal forcing.
\rfig{p2} also includes data from a large number of 
studies which include 
both solenoidally and dilatationally forced isotropic turbulence  
as well as shear flows. These are summarized
in \rtab{sources}.
It is also clear that the scaling with $\mt$ proposed in \cite{DJ2013} 
for solenoidal forcing (dashed line) is 
inadequate for the general case. 
As we will see momentarily universal scaling 
do in fact emerge when one uses 
an appropriate non-dimensional group which incorporates dilatational
motions.

Our data in \rfig{p2}a come from highly resolved 
direct numerical simulations (DNS) of 
the compressible Navier-Stokes equations representing
conservation of mass, momentum and energy. To provide
a full set of closed equations we use, as commonly 
done, a perfect gas to relate thermodynamic 
quantities, and molecular transport terms 
with a power-law dependence on temperature.
The simulations use tenth-order compact schemes
for spatial differentiation and third order Runge-Kutta 
in time. The momentum equations are forced 
stochastically at the largest scales (low 
wavenumbers) using independent 
Ornstein-Uhlenbeck random processes with 
finite time correlation. In Fourier space,
one can project the Fourier modes on a plane 
perpendicular to the
wavenumber vector or parallel to it, resulting in 
solenoidal and dilatational forcing, respectively.
A statistically stationary state is maintained 
by removing 
energy from the system uniformly through the energy equation.
We find the results here to be insensitive of whether energy 
is removed or not or the details of energy
removal.
Further details of the numerical scheme as well as 
detailed statistics of the resulting flow 
can be found in \cite{DJ2013,JD2016}.

\tikzstyle{everypath}=[line width=1pt]

%%%%%%%%%%%%% start of table 2 %%%%%%%%%%%%%%%%%%%%%%%%%%%%
%\captionsetup[table]{position=below}   %% or below
\begin{table}
\begin{tabular}{|c|c|c|c|c|}
\hline
\fs{Flow type}  & \fs{$\sigma$}& \fs{$\mt$} & \fs{$\re$} & \fs{Symbol} \\
%\hline
%\fs{HIT, S} \cite{JD2016} & \fs{100} & \fs{0.1-0.6} & \fs{32-430} & \tikz [line width=1pt] \draw[mycoljd] (1,0.75) circle (0.1); \\
\hline
\fs{HIT, S}  \cite{WGWPRF2017} & \fs{100} & \fs{0.05-1.02}  & \fs{38-370} & \tikz [line width=1pt] \draw[mycolgot] (1,0.75) circle (0.1); \\
\hline
\fs{HIT, S} \cite{KO1990,KO1992} & \fs{100} & \fs{0.11-0.88}  & \fs{12-44}  &\tikz [line width=1pt] \draw[mycolkidas] (1,0.75) circle (0.1); \\
\hline
\fs{HIT, S}  \cite{WSWXHCPRL2012,WSW+2011} & \fs{100} & \fs{1.03} & \fs{254}  &\tikz [line width=1pt]  \draw[mycolwangprl2012] (1,0.75) circle (0.1); \\
\hline
\fs{HIT, S (Present)} & \fs{100} & \fs{0.05-0.6} & \fs{38-430} $\dagger$ &
\tikz \draw[mycolwanghst2018,fill=mycolwanghst2018] (1,0.75) circle (0.075);  
\tikz \draw[mycolgot,fill=mycolgot] (1,0.75) circle (0.075);  
\tikz \draw[mycolwangjt,fill=mycolwangjt] (1,0.75) circle (0.075); 
\tikz \draw[mycolwangprl2012,fill=mycolwangprl2012] (1,0.75) circle (0.075); 
\tikz \draw[mycolwangpre,fill=mycolwangpre] (1,0.75) circle (0.075); 
\tikz \draw[mycolkidas,fill=mycolkidas] (1,0.75) circle (0.075);
\\
%\tikz \draw[black,fill=black] (1,0.75) circle (0.1); \\
\hline
\fs{HIT, D} \cite{KO1990,KO1992} & \fs{0} & \fs{0.11,0.17} & \fs{5.4,10}& \tikz [line width=1pt] \path[draw, color=mycolkidad] (1.6,0.75)--(1.5,0.9)--(1.7,0.9)--cycle ; \\
\hline
\fs{HIT, D} \cite{WYSXHC2013} & \fs{33.33} & \fs{0.73} & \fs{210} & \tikz [line width=1pt] \path[draw, color=mycolwangjt] (1.6,0.75)--(1.5,0.9)--(1.7,0.9)--cycle ; \\
\hline
\fs{HIT, D} \cite{WJWMCSXCCS2018} & \fs{50} & \fs{0.30-0.65} & \fs{196-234} &\tikz [line width=1pt] \path[draw, color=mycolwangpre] (1.5,0.75)--(1.6,0.9)--(1.7,0.75)--cycle ; \\
\hline
%\fs{HIT, D} \cite{WYSXHCPRL2013} & \fs{50??} & \fs{0.62} & \fs{160} & \tikz [line width=1pt] \path[draw, color=mycolwangprl2013] (1.5,0.75)--(1.6,0.9)--(1.7,0.75)--cycle ; \\
%\hline
\fs{HIT, D} \cite{ni2016} & \fs{4.76, 50}  $\dagger$ & \fs{0.6} & \fs{180} &
\tikz [line width=1pt] \path[draw, color=mycolsar1] (1.6,0.75)--(1.5,0.9)--(1.7,0.9)--cycle ; 
\tikz [line width=1pt] \path[draw, color=mycolsar1] (1.5,0.75)--(1.6,0.9)--(1.7,0.75)--cycle ;
\\
\hline
\fs{HIT, D (Present)}  & \fs{70-90} & \fs{0.04-0.8}  & \fs{30-160} $\dagger$ &
\tikz \draw[mycolwanghst2018,fill=mycolwanghst2018] (1.5,1.1)--(1.5,0.9)--(1.65,1.0)--cycle ; 
\tikz \draw[mycolgot,fill=mycolgot] (1.5,1.1)--(1.5,0.9)--(1.65,1.0)--cycle ;  
\tikz \draw[mycolwangjt,fill=mycolwangjt] (1.5,1.1)--(1.5,0.9)--(1.65,1.0)--cycle ; 
\tikz \draw[mycolwangprl2012,fill=mycolwangprl2012] (1.5,1.1)--(1.5,0.9)--(1.65,1.0)--cycle ; \\
\hline
\fs{HIT, D (Present)}  & \fs{35-70} & \fs{0.05-0.55}  & \fs{38-154} $\dagger$  &
%\tikz \path[draw, color=blue,fill=blue] (1.5,1.1)--(1.5,0.9)--(1.65,1.0)--cycle ;, 
%\tikz \path[draw, color=magenta,fill=magenta] (1.5,1.1)--(1.5,0.9)--(1.65,1.0)--cycle ;, 
%\tikz \path[draw, color=green,fill=green] (1.5,1.1)--(1.5,0.9)--(1.65,1.0)--cycle ;, 
%\tikz \path[draw, color=red,fill=red] (1.5,1.1)--(1.5,0.9)--(1.65,1.0)--cycle ; \\ 
\tikz \draw[mycolwanghst2018,fill=mycolwanghst2018] (1.5,0.75)--(1.6,0.9)--(1.7,0.75)--cycle ; 
\tikz \draw[mycolgot,fill=mycolgot] (1.5,0.75)--(1.6,0.9)--(1.7,0.75)--cycle ;  
\tikz \draw[mycolwangjt,fill=mycolwangjt] (1.5,0.75)--(1.6,0.9)--(1.7,0.75)--cycle ; 
\tikz \draw[mycolwangprl2012,fill=mycolwangprl2012] (1.5,0.75)--(1.6,0.9)--(1.7,0.75)--cycle ; \\
\hline
\fs{HIT, D (Present)}  & \fs{0-35} & \fs{0.04-0.25}  & \fs{16-77} $\dagger$ &
%\tikz \path[draw, color=blue,fill=blue] (1.5,1.1)--(1.5,0.9)--(1.65,1.0)--cycle ;, 
%\tikz \path[draw, color=magenta,fill=magenta] (1.5,1.1)--(1.5,0.9)--(1.65,1.0)--cycle ;, 
%\tikz \path[draw, color=green,fill=green] (1.5,1.1)--(1.5,0.9)--(1.65,1.0)--cycle ;, 
%\tikz \path[draw, color=red,fill=red] (1.5,1.1)--(1.5,0.9)--(1.65,1.0)--cycle ; \\ 
\tikz \draw[mycolwanghst2018,fill=mycolwanghst2018] (1.6,0.75)--(1.5,0.9)--(1.7,0.9)--cycle ; 
\tikz \draw[mycolgot,fill=mycolgot] (1.6,0.75)--(1.5,0.9)--(1.7,0.9)--cycle ;  
\tikz \draw[mycolwangjt,fill=mycolwangjt] (1.6,0.75)--(1.5,0.9)--(1.7,0.9)--cycle ; \\
\hline
%\tikz \path[draw, color=black,fill=black] (1.5,1.1)--(1.5,0.9)--(1.65,1.0)--cycle ; \\
\fs{HST}  \cite{CWLWC2018} & \fs{...}  & \fs{0.03-0.66}  & \fs{32-220}&\tikz [line width=1pt] \path[draw, color=mycolwanghst2018] (1.5,1.05)--(1.5,0.9)--(1.65,0.9)--(1.65,1.05)--cycle ; \\
\hline
\fs{HST}  \cite{SEHYTCFD1991}$^{*}$   & \fs{...}  & \fs{0.2-0.5} & \fs{16-35} &  \tikz  [line width=1pt] \path[draw, color=mycolgot] (1.5,1.05)--(1.5,0.9)--(1.65,0.9)--(1.65,1.05)--cycle ; \\
\hline
\fs{HST}  \cite{sarkar1992compressible} & \fs{...}  & \fs{0.2-0.7} $\dagger$  & \fs{14-45} &\tikz  [line width=1pt] \path[draw, color=mycolwangjt] (1.5,1.05)--(1.5,0.9)--(1.65,0.9)--(1.65,1.05)--cycle ; 
\tikz  [line width=1pt] \path[draw, color=mycolkidas] (1.5,1.05)--(1.5,0.9)--(1.65,0.9)--(1.65,1.05)--cycle; \\
\hline
\fs{HST}  \cite{sarkarjfm1995}$^{*}$ & \fs{...}  & \fs{0.13-0.65} & \fs{14-32} &\tikz  [line width=1pt] \path[draw, color=mycolwangprl2012] (1.5,1.05)--(1.5,0.9)--(1.65,0.9)--(1.65,1.05)--cycle ; \\
\hline
\fs{HIT, TF}  \cite{CGTFJFM2019} & \fs{...}  & \fs{0.2,0.6}  & \fs{250}&\tikz  [line width=1pt] \path[draw, color=mycolwangtherm] (1.55,0.75)--(1.45,0.65)--(1.55,0.55)--(1.65,0.65)--cycle ; \\
\hline
%\fs{HIT} \cite{} &\fs{0,50,5}
\end{tabular}
\caption{
\label{tab:sources} 
Databases used in the present study.
Flow types:
homogeneous isotropic turbulence (HIT),
homogeneous shear turbulence (HST, squares),
which can have 
solenoidal forcing (S, circles), 
some dilatational forcing (D, triangles),
or thermal forcing (TF). 
Studies with multiple symbols correspond to the different
conditions marked with a $\dagger$ in increasing order.
Studies marked with an asterisk did not provide 
$\delta$ and was thus computed using pressure fluctuations
and \req{p2_dim}.
%
%forcing, circles),
%HIT, D (homogeneous isotropic tubulence, dilatational forcing, triangles 
%depending on $\sigma$), and HST (homogeneous shear, squares)
% and TF (thermal forcing, rhombus). 
%Color schemes of present data based on 
%$R_{\lambda}$ with  \textcolor{mycolwanghst2018}{$R_{\lambda} < 40$}, 
% \textcolor{mycolgot}{$40 < R_{\lambda} < 70$}, 
% \textcolor{mycolwangjt}{$70 <R_{\lambda} < 115$}, 
% \textcolor{mycolwangprl2012}{$115 < R_{\lambda} < 180$}, 
% \textcolor{mycolwangpre}{$180 < R_{\lambda} < 300$}, 
% \textcolor{mycolkidas}{$300 < R_{\lambda} < 500$}. 
%The symbols and 
%color schemes used in the table will be used for all subsequent
%figures.
 }
% \begin{picture}(0,0)
% \put(-120,-8){\tiny * $\delta$ was not available directly, was derived by assuming scaling of pressure ratio we derived in this paper.}
% %\put(-120,-14){\tiny + Different color corresponds to different initial turbuelnt Mach number 
% %with \textcolor{mycolwangjt}{$M_{t0}$:0.2} and \textcolor{mycolkidas}{$M_{t0}:0.4$} }
% % %\put(-235,-93){\tiny The derived $\delta$ was used for scaling of other statistics, ** $\delta$ is based of transverese velocity components}
% \end{picture}
\end{table}

%%%%%%%%%%%%%%%%%%%%%%%%%%%%%%%%%%%%%%%end of table 2 %%%%%%%%%%%%%%%%%%%%%%%%%%%%%%%%%%

\section{Self similarity for compressible flows}

Returning to the discussion on the general principles around scaling 
in incompressible turbulence, we are thus confronted with the same 
difficulties but on a larger parameter space. 
As argued above,
one needs to identify a dilatational velocity 
${\cal U}_d$ that characterize the content of dilatational 
motions. Unfortunately, it is unclear how this velocity 
can be obtained from geometrical aspects of a given flow
such as the geometry of a grid.
And even if it was possible for one particular flow, the results 
would typically depend on the details of the setup, as we
illustrate below.
What we propose instead is to rely on the same approach 
used in incompressible turbulence, namely, 
to seek internal similarity in which we use an internal 
scale generated by the flow. 
A natural choice would be 
$\udrms$, the root-mean-square of the 
dilatational component of velocity based on the
Helmholtz decomposition.
%\footnote{It is also possible to 
%apply Helmholtz decomposition to $\sqrt{\rho}\vec{u}$ \cite{KO1990}.
%Results here are not sensitive to this choice, though.}. 
This is an analogous reasoning that leads 
to the selection of $\urms$ for incompressible flows.
The parameter space is now augmented such that
\be
Q=f(L, \mu, \la\rho\ra, \urms, c, \udrms).
\label{eq:parameters}
\ee
Note that only a single length scale is included 
here which can be justified in the grounds 
that the geometry of the device, for example, will 
provide a natural length scale for the problem
with no distinction between solenoidal and dilatational
components. The validity of such an assumption 
will ultimately rest on its success in providing universal
scaling laws to numerical and experimental data.

Dimensional analysis would then reduce the
list of parameters from six to three.
Different non-dimensional groups can be formed
containing dilatational motions. 
Three such parameters are 
$\delta=\udrms/\usrms$,
$\chi\equiv K_d/K$
(where $K_d=\la \rho|\ud|^2\ra/2$ and 
$K=\la\rho|\ut|^2\ra/2$ are 
the turbulent kinetic energy in the dilatational mode
and both modes combined, respectively),
and
$\mtd \equiv \udrms/c$ 
which compares a typical velocity of dilatational
motions to that of acoustic propagation.
The parameter $\chi$ is perhaps the most widely used,
while $\mtd$ has been explored in astrophysical 
contexts \cite{KGFK2012}.
It is trivial to find relationships such as 
$\chi \approx \delta^2/(\delta^2+1)$
or $\mtd\approx \mt\sqrt{\chi}$
where the approximations stem for neglecting density 
correlations, which have verified 
%We have verified 
%these minor differences are not of concern in the 
%objectives of this work.
to be minor for the objectives of this work.
%Two such parameters are 
%%${\Upsilon}=\udrms/\urms$
%and $\delta=\udrms/\usrms$.
%These are related to another parameter
%widely used in the literature, namely 
%$\chi\equiv K_d/K$
%where $K_d=\la \rho|\ud|^2\ra/2$ and 
%$K=\la\rho|\ut|^2\ra/2$ are 
%the turbulent kinetic energy in the dilatational mode
%and both modes combined respectively (note that because
%of the orthogonality of Helmholtz modes for 
%the velocity field, one sees $K\approx K_s + K_d$
%with $K_s=\la\rho|\us|^2\ra/2$ due to the relatively 
%weak correlation observed with density).
%It is trivial to find relationships between 
%these parameters such as 
%$\chi\approx \Upsilon^2\approx \delta^2/(\delta^2+1)$
%where the approximation stems for neglecting density 
%correlations.
%%For a range of conditions,
%%one can approximate $\Upsilon\approx\sqrt{\chi}$
%%where the approximation stems from 
%%$\chi$ containing the fluctuating density
%%in its definition.
%We have verified that 
%these minor differences are not of concern in the 
%objectives of this work.
%%Similarly one can show $\delta \approx  \sqrt{\chi/(1-\chi)}$
%%or $\chi\approx \delta^2/(\delta^2+1)$.
Note that searching for the scaling of these parameters
with $\mt$, which has been the focus of some investigations, 
implicitly assumes them to be dependent 
parameters of the problem. For example, 
$\chi\sim \mt^2$ or $\mt^4$ 
have been suggested based on 
different assumptions on EDQNM closures 
\cite{SC.book2008}. 
However, as we have seen in \rfig{p2}a, 
$\mt$ and $\re$ alone cannot describe completely 
the statistical state of turbulence in the 
general case.
%Another parameter that can be formed, is the so-called 
%dilatational Mach number as $\mtd \equiv \udrms/c$ 
%which compares a typical velocity of dilatational
%motions to that of acoustic propagation.
%This parameter has been explored in astrophysical 
%contexts \cite{KGFK2012}.
%The two parameters are related 
%according to $\mtd = \Upsilon\mt \approx \sqrt{\chi}\mt$.

Our main objective here 
is to explore the use of two non-dimensional 
groups such as $\mt$ and $\delta$ 
as the proper similarity parameters.
Formally, we suggest that self-similar scaling is 
possible in compressible flows if internal dilatational
scales are included to form similarity parameters. That is, 
we propose 
\be
%\overline{Q}=f_c(\re,\mtd,\chi) .
\overline{Q}=f_c(\re,\mt,\delta) .
\label{eq:css_new}
\ee
instead of \req{css} as a base of finding universality in 
such flows. While other non-dimensional parameters 
can be formed, as discussed above, 
$\mt$ has been used extensively as a governing parameter 
and is used here complemented by $\delta$.

The addition of $\delta$ 
as a parameter to characterize compressibility 
highlights another important aspect that may not be 
immediately apparent.
The turbulent Mach number $\mt$ has been traditionally used to 
characterize two aspects of compressible turbulence.
On the one hand, it has been used as a measure of 
compressibility levels, which can be interpreted as 
the strength of dilatational motions 
%(as oposed to solenoidal,
%incompressible motions) 
in the flow.
On the other hand,
$\mt$ has also been widely interpreted as the ratio of acoustic 
to turbulence time scales. This follows from 
using the so-called eddy-turnover time, 
$L/\urms$,
to characterize the large turbulent scales and identifying 
the acoustic time with $L/c$ which represents the 
time for an acoustic wave to transverse a distance of 
the order of the largest scales in the flow. 
The ratio is obviously $\mt$ and measures the 
disparity at which acoustic and turbulence processes
occur. A large time disparity (low $\mt$)
was assumed to obtain the linear 
system in \req{linear} \cite{EHKS1990}.
The dual role traditionally assigned to $\mt$, thus, 
conceptually forbids situations where 
acoustic phenomena occur at short time scales (low $\mt$)
but strong compressibility effects (high $\delta$) 
are present. 
%This situation, characterized by low $\mt$
%and relatively high $\delta$
%cannot be captured if $\mt$ 
%is a measure of both.
The alternative use of $\mtd$, 
$\chi$ or $\delta$ to characterize compressibility 
levels, thus, does not restrict 
\req{linear} to flows with small dilatational content. 

This may help explain the 
accuracy and robustness of equipartition 
($F\approx 1$ as shown below) for flows with relatively 
large values of $\mt$ \cite{SEHK1991}, 
which is unexpected given that 
equipartition is based on \req{linear}, assumed to 
be valid only in the low $\mt$ limit.
However, in \cite{SEHK1991}, as the initial $\mt$ was increased,
so was $\chi$. It is, thus, possible
that at high $\chi$, dilatational terms in the governing
equations may dominate 
the dynamics even though scale separation was not that 
large (relatively high $\mt$). 
Therefore, the linear system
\req{linear} may be valid for a wider range of 
conditions than originally thought.
 
%In fact, it is also expected that since 
%dilatational motions become more dominant
%as $\mtd$ is increased, the 
%time disparity between modes may not need to be so great
%(i.e.\ $\mt$ does not need to be so small) for the simplified 
%system \req{linear} to provide a reasonable approximation of 
%the dilatational structure the flow.
%This may be so, in part, because the neglected terms in the system 
%of equations because of time disparity may now
%be relatively smaller when strong dilatational motions 
%are externally imposed.
%This may help explain the 
%accuracy and rebustness of equipartition ($F\approx 1$ as shown below) 
%for flows with relatively 
%large values of $\mt$ as observed in 
%\cite{SEHK1991} where data was presented 
%for initial conditions with higher $\chi$ 
%as $\mt$ was increased.

%In the next section
%we resort to asymptotic equilibrium states
%to construct scaling laws from which 
%non-trivial governing
%parameters---that is, parameters that cannot be 
%derived from dimensional considerations alone---naturally emerge.
%We then use DNS data
%to test the scaling in \req{css_new}.

\section{The scaling of pressure in compressible turbulence}

In \rfig{p2}a we showed the dramatic
effect of dilatational forcing on the variance 
of pressure.
We have also argued,
with \req{continuity} and \req{linear}, that
fluctuations of thermodynamic variables
are more closely related to dilatational motions,
and proposed \req{css_new}.
Here we will look at the specific case 
of pressure fluctuations for which known statistical 
equilibrium states can be used 
to justify more rigorously this proposal.

Consider the pressure variance, which 
according to \req{css_new} can be written as 
$\prms^2/\la p\ra^2=f_c(\re,\mt,\delta)$
where the mean pressure
is formed with the governing parameters 
as $\la p\ra = c^2\la\rho\ra/\gamma$.
In the incompressible limit
it is well-established that
pressure finds a statistical equilibrium with 
$\prms\approx A\rhom \usrms^2$
with a very weak Reynolds number dependence 
\cite{LL.book.1987,DSY2012}.
Using definitions, this expression can be 
re-written as 
\begin{equation}
\prms^2/\la p\ra^2 \approx
	A^2 \gamma^2 \mt^4/9 .
\label{eq:p2s}
\end{equation}
This scaling has been verified with solenoidally forced simulations 
in \cite{DJ2013,WGWPRF2017} also seen in
 \rfig{p2}a (circles and dashed line).
%Shear flows (red squares)\cite{CWLWC2018} also follow an approximate 
%$\mt^4$ scaling though, as expected,
%with a slightly different constant $A$.

Now consider the purely dilatational case governed by 
the system \req{linear}.
It is easy 
to obtain an analytical solution for $u_d$ and $p_d$.
For any initial condition, one can show that for unforced
flows the solution tends to an equilibrium state 
of equipartition between dilatational kinetic 
energy and potential
energy in the dilatational pressure \cite{SEHK1991}. 
While in the literature this is simply called
equipartition, here we will call it $p$-equipartition to 
distinguish it from another form of equipartition 
discussed below.
Explicitly, this statistical state can be 
expressed as $F_p=1$ using the 
so-called equipartition function $F_p\equiv c_0^2\rho_0^2
\udrms^2/\pdrms^2$.
While this result involves only the dilatational 
pressure, 
Ref.~\cite{JD2016} shows that 
as compressibility levels increase, 
the dilatational pressure becomes more dominant.
This observation implies that at high 
compressibility levels (a condition
to be defined momentarily), the dilatational
pressure would be indeed a proxy to assess the 
scaling of total pressure.
In this case, 
which we term dilatationally dominated $p$-equipartition,
or \DDE, we would then
expect $\prms^2 = c_0^2\rho_0^2\udrms^2/F_p$
or in non-dimensional form
\be
\prms^2/\la p\ra^2 = \gamma^2 \mtd^2 /F_p \ .
\label{eq:p2}
\ee
Clearly this scaling does not conform with \req{css}
but does agree with the proposed \req{css_new}
since 
$\mtd\approx \mt \delta/\sqrt{\delta^2+1}$.
%Indeed, as seen in \rfig{p2}b, 
%\req{p2} shows excellent agreement with DNS 
%data at high $\mtd$.
In \rfig{p2}b we test the scaling with $\mtd$. 
Comparision between parts (a) and (b) reveals a
much better
collapse of the data on a universal curve 
at high $\mtd$.
Since the database contains cases where dilatational
motions are directly driven by an external dilatational 
forcing and also cases where
dilatational motions appear exclusively
due to Navier-Stokes dynamics (solenoidal forcing or 
homogeneous shear flows),
the collapse in the figure supports the idea of 
universal scaling when an internal 
dilatational velocity scale is used, regardless of 
the means of generation. Reynolds number 
effects are found to be negligible 
consistent with previous studies \cite{DJ2013,JD2016}

These results have clear implications for the scaling of
compressible turbulence. 
$P$-equipartition, as expressed in \req{p2}, is a
theoretical result that follows from neglecting several terms,
including those representing viscous and non-linear processes,
in the governing equations \cite{SEHK1991}. 
%A more general justification of equipartition between potential
%and kinetic energy can also be formulated using classical 
%statistical mechanical methods, in particular, by the Liouville's
%theorem in a phase space constructed with the Fourier modes 
%of the velocity field and a 
%specific bijective function of density \cite{kraich1955}.
The fact that $p$-equipartition
is indeed observed at {\it some}
condition (which is identified below), forces us to conclude that 
\req{css} is fundamentally deficient in the general case.

By construction we have $p=\ps+\pd$ which implies
$\prms^2 = \psrms^2 + \pdrms^2 + 2 r \, \psrms\pdrms$
where $r$ is the correlation coefficient between 
$\ps$ and $\pd$. 
In the two asymptotic cases discussed above, however, 
one component of pressure will be much larger than the other:
either $\prms \approx \psrms$ or $\prms\approx \pdrms$.
This will also be the case if the two components 
of pressure are only weakly correlated as was
found at low $\mt$ \cite{JD2016}. 
We thus have
\be
\frac{\prms^2}{\pm^2} \approx 
A^2\gamma^2\frac{\mtd^4}{\delta^4} + 
\frac{\gamma^2}{F}\mtd^2
\label{eq:p2_dim}
\ee
or, for convenience, one can re-write it in 
non-dimensional form as
\begin{equation}
{\prms^2 \over \pm^2} {F\over \mtd^2\gamma^2} 
   \approx  A^2F\D^{-2} + 1
\label{eq:p2_D}
\end{equation}
where the new parameter
\begin{equation}
\D \equiv {\delta \sqrt{\delta^2+1}/\mt} 
\label{eq:D}
\end{equation}
is a measure of the relative dominance of the dilatational 
to the solenoidal contributions.
We can also write, in terms of different parameters,
$\D = 
   {\delta^2}/{\mtd} 
    = {\chi/\mtd (1-\chi)} 
    = {\sqrt{\chi}/\mt (1-\chi)}
    $.
Note that $\D$ is a complex
combination of the originally proposed non-dimensional
groups in \req{css_new} and can thus be considered a
consequence of 
self-similarity of the second kind \cite{BZ1972} in which 
governing non-dimensional groups cannot be obtained by 
dimensional analysis alone. 

%, that is 
%$\pdrms^2/\psrms^2=A^2F\D^{-2}$ where, as mentioned above
%$F$ and $A$ are $O(1)$ constants.

This relation presents some interesting consequences.
First, it suggests $\D$ as an appropriate parameter
to determine not only levels of pressure fluctuations 
but also, the statistical regime one expects turbulence
to be in. 
At high $\D$, dilatational pressure dominates and 
$p$-equipartition is the main
mechanism governing the dynamics of pressure fluctuations. 
%This condition can be termed dilatationally-dominated 
%equiparition.
At low $\D$, pressure is dominated by its elliptic 
nature dictated by the incompressible Navier-Stokes equations.
In $p$-equipartition one has $F=1$, and since $A\sim O(1)$ for all 
flows, one would expect a critical value $\D=\D_{crit}$ of order
unity that separates the two regimes. 
The specific value of $\D_{crit}$ 
as well as the asymptote at high $\D$, however, will retain
a (weak) dependence on flow characteristics through 
the constant $A$.

These observations are indeed consistent with 
\rfig{p2}c where we show the data normalized according
to \req{p2_D}. There is excellent collapse of the data
along these two asymptotic equilibrium states with a 
relatively sharp transition around $D_{crit}\approx 0.5$. 
This transition towards a flow dominated by dilatational 
pressure relative to solenoidal pressure 
was also observed in Ref.~\cite{JD2016} (also included here)
though the use of $\mt$ instead of $\D$
(perhaps justified since forcing was solenoidal and identical
in nature for all runs) results in larger scatter. 
%Those simulations, 
%which are also included here, are seen to be consistent with the rest
%of the data.

\rfig{p2}c also includes homogeneous shear flows 
from Refs.~\cite{CWLWC2018, SEHYTCFD1991,sarkar1992compressible,sarkarjfm1995} where no explicit 
forcing term is added to the governing equations.
Instead, turbulent fluctuations  
are generated as a result of the production mechanism 
by the mean shear. The good collapse with other data provides
further support for universal 
self-similar scaling for diverse flows driven by different 
mechanisms. 

We do point out, though, that while 
the constant $A$ in \req{p2s} is order unity, its 
numerical value is flow dependent. In fact, for the shear flow
studies we find $A\approx 1.9$ which is slightly different 
than 1.2 in stochastically forced isotropic flows \cite{JD2016}.
Different low-$\D$ asymptotes are therefore observed for
these cases. On the other hand, if full $p$-equipartition 
governs the flow ($F=1$), then the asymptotic state is 
universal.

To put forth \req{p2_D} we have assumed that the solenoidal 
pressure behaves as in incompressible flows and 
that the dilatational part behaves according to $p$-equipartition 
for {\it all} conditions. This would imply that, 
for example, $\psrms=A\rhom\usrms^2$ regardless of the 
amount of dilatation in the flow. However, 
weaker assumptions are actually needed, namely, that the 
solenoidal pressure is not affected by the dilatational 
component when solenoidal pressure dominates the 
total pressure. Thus, the success of \req{p2_D}
to collapse the data does not preclude other behavior for $\psrms$
when dilatational pressure dominates.
The same can be said about dilatational pressure, namely, 
that it is not affected by the solenoidal part only when 
dilatational pressure dominates and that other scaling 
behavior is possible under those conditions. 

Another interesting observation from \req{D} is the 
dependence on $\mt$. While throughout the literature this is 
considered to be a measure the level of compressibility,
\req{D} seems to suggest that, an increase 
in $\mt$ (at constant $\delta$)
leads to a decrease in $\D$ which would 
represent weaker compressibility effects. 
%In real flows, however, this may not be the case
%as both $\mt$ and $\delta$ may change simultaneously. 
This result and its broader implications will be discussed in
the Discussion section. %\rsec{broad}.

We finally note that in a number of circumstances low-order
statistics of thermodynamic variables can be related 
to a good approximation by an isentropic, or more generally, 
polytropic process \cite{JD2016, DJ2013,WGWPRF2017,WJWMCSXCCS2018,CWLWC2018,WGWPF2017inter}. 
In this case, one finds $\Trms/\Tm$ and
$\rhorms/\rhom$ to exhibit the same universal scaling seen 
in \rfig{p2}c with slightly different prefactors.
%The interesting conclusion from this analysis is, thus, 
%that there is indeed self-similar scaling describing the 
%statistical thermodynamic state of compressible turbulence 
%which appears to be universal. 
% %However, unlike
% %many of the problems mentioned in the introduction 
% %where analogies with critical phenomena have proven
% %successful in describing universal aspects of the 
% %macroscopic system, we show below that 
% %multiple transitions occur in a larger parameter 
% %space. 
% %Still, we will argue that self-similar universal 
% %scaling is possible on this larger parameter space.

\begin{figure}
\begin{center}
\includegraphics[clip,trim=220 00 220 00,width=0.22\textwidth]{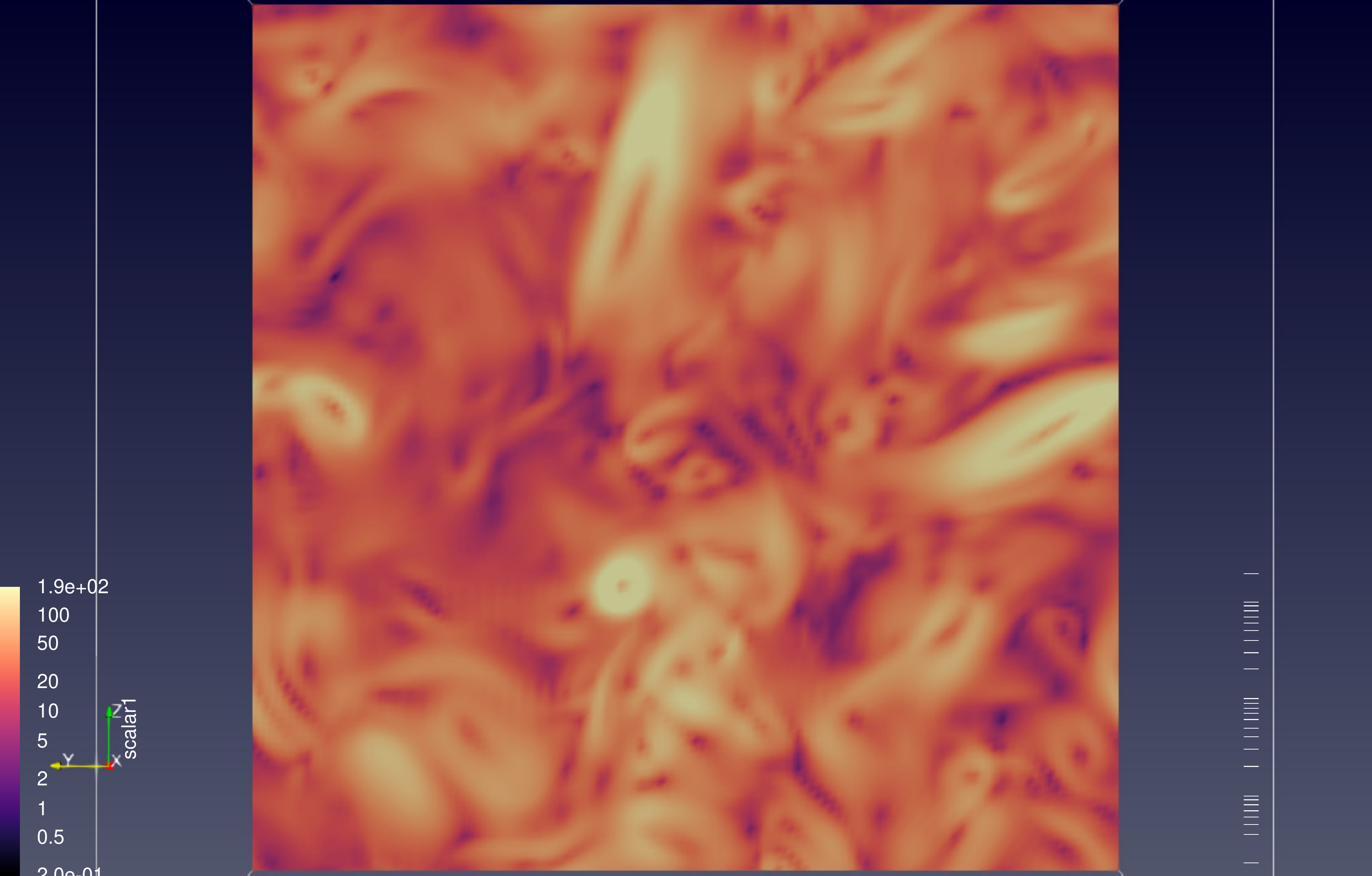}
\includegraphics[clip,trim=220 00 220 00,width=0.22\textwidth]{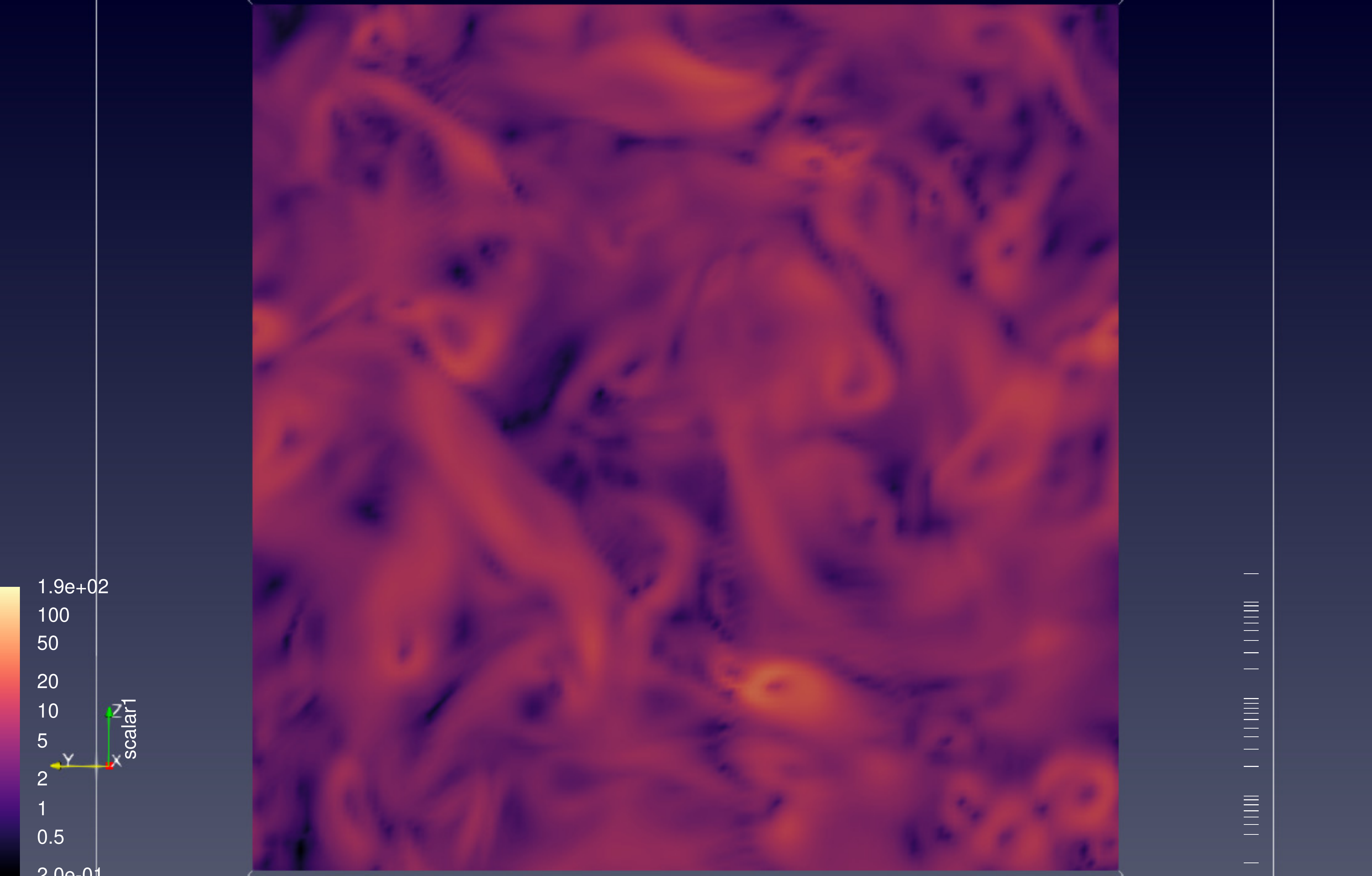} 
\includegraphics[clip,trim=220 00 220 00,width=0.22\textwidth]{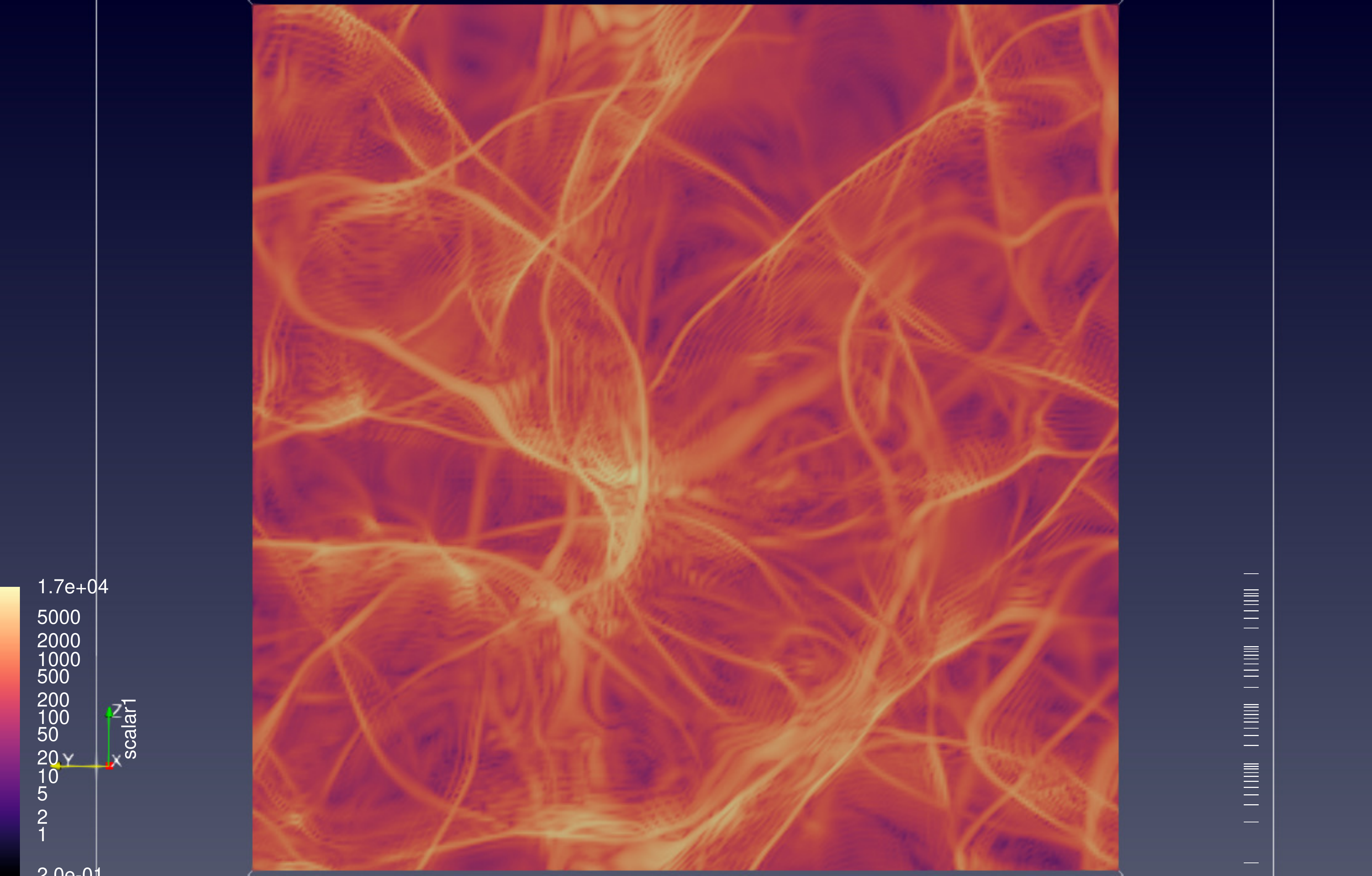}
\includegraphics[clip,trim=220 00 220 65,width=0.22\textwidth]{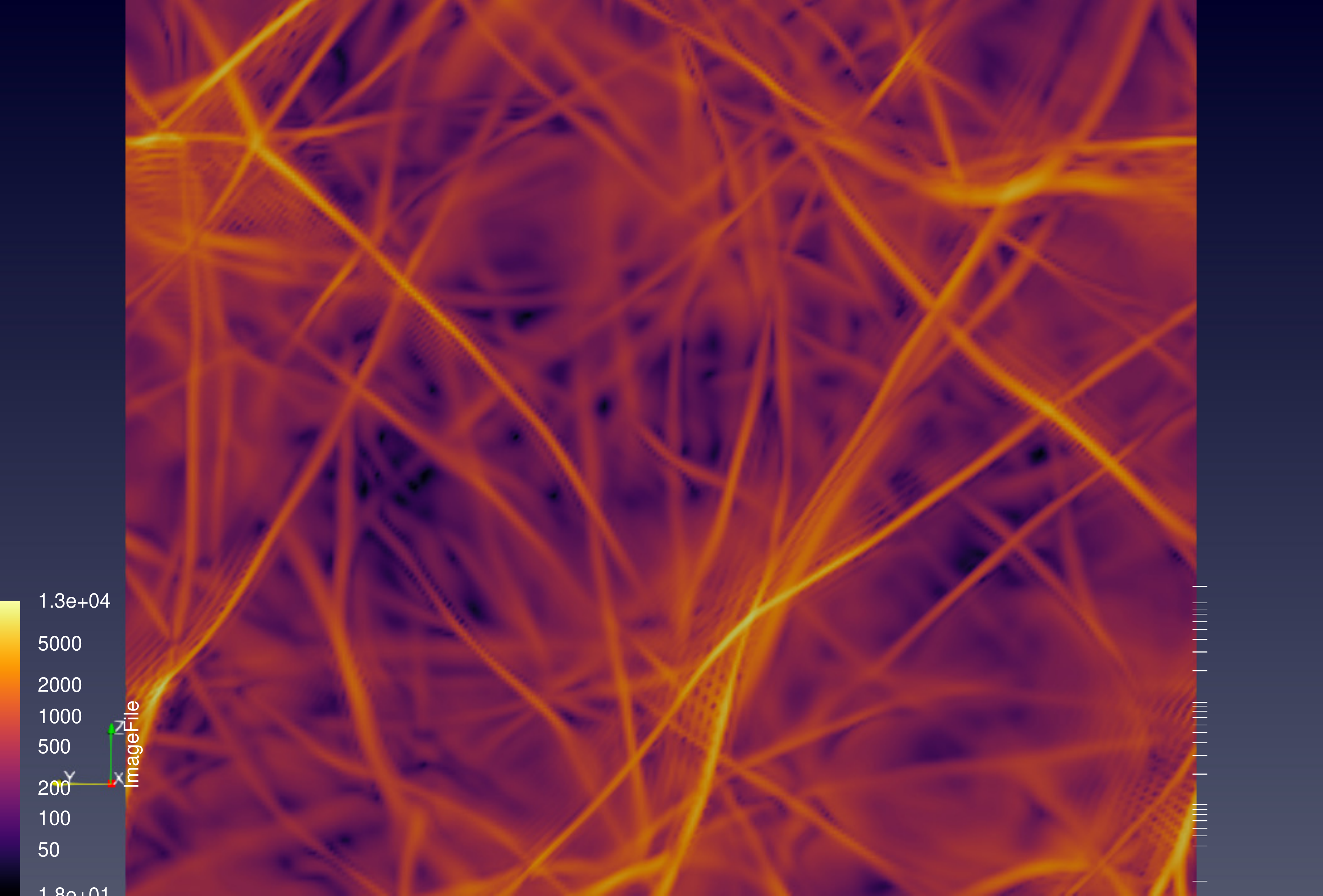}
\begin{picture}(0,0)
%\put(-67,155){\small{ $M_{t}$: 0.09,$R_{\lambda}$: 58,$\delta$: 0.0035$,$D$: 0.0388}}
%\put(-85,140){\small $M_{t}: 0.09$}
%\put(-45,140){\small $R_{\lambda}: 58$}
%\put(-85,125){\small $\delta: 0.0035$}
%\put(-45,125){\small $D: 3.9E^{-02} $}
%%%%%%%%%%%%%%%%%%%%%%%%%%%%%%%%%%%%%%%%%%
%\put(25,140){\small $M_{t}: 0.36$}
%\put(65,140){\small $R_{\lambda}: 102$}
%\put(25,125){\small $\delta: 0.097$}
%\put(65,125){\small $D: 2.7E^{-01} $}
%%%%%%%%%%%%%%%%%%%%%%%%%%%%%%%%%%%%%%%%%
%\put(25,25){\small $M_{t}: 0.05$}
%\put(65,25){\small $R_{\lambda}: 33$}
%\put(25,10){\small $\delta: 5.447$}
%\put(65,10){\small $D: 5.6E^{02} $}
%%%%%%%%%%%%%%%%%%%%%%%%%%%%%%%%%%%
%\put(-85,25){\small $M_{t}: 0.20$}
%\put(-45,25){\small $R_{\lambda}: 50$}
%\put(-85,10){\small $\delta: 1.40$}
%\put(-45,10){\small $D: 1.2E^{01} $}
%%%%%%%%%%%%%%%%%%%%%%%%%%%%%%%%%%%%%%%%
\end{picture}
\end{center}
\caption{
\label{fig:viz}
Contours of $\left| \nabla p\right|$ for 
$({\cal D},\re,\mt,\delta) = (0.04,58,0.09,0.0035)$ (a),
$(0.27,102,0.36,0.097)$ (b),
$(12,50,0.20,1.4)$ (c),
and
$(52,51,0.06,1.65)$ (d).
%a) ${\cal D}= 0.04$, $M_{t}= 0.09$, $\delta= 0.0035$ and $R_{\lambda}= 58$
%b) ${\cal D}= 0.27$, $M_{t}= 0.36$, $\delta= 0.097$ and $R_{\lambda}= 102$
%c) ${\cal D}= 12$, $M_{t}= 0.20$, $\delta= 1.400$ and $R_{\lambda}= 50$
%d) ${\cal D}= 566$, $M_{t}= 0.05$, $\delta= 5.447$ and $R_{\lambda}= 33$
}
\end{figure}

The spatial structure of the flows in the different regimes
is also qualitatively different.
In \rfig{viz} we show the instantaneous pressure gradient at an 
arbitrary plane and instant of time.
The flow is seen to exhibit completely different features even 
when $M_{t}$ is similar (a and d) showing again the inadequacy 
of $\mt$ to capture important aspects of the flow.
In (a) and (b),  (${\cal D} < {\cal D}_{crit}$)
we observe that 
pressure gradient contours look similar to those seen in incompressible 
turbulence, consistent with the fact that the pressure field is 
dominated by the solenoidal pressure.
Beyond the transitional $D_{crit}\approx 0.5$ (panels c and d)
we observe 
very thin high pressure gradient fronts resembling shock waves 
in contrast to the more isotropic vortical high and low gradient 
regions observed  for cases with $D < D_{crit}$. 
However, some differences are also apparent between panels c and d.
The shock-like structures in c appear more curved than those
in d indicating 
perhaps a stronger coupling between the vortical solenoidal motions
and the strong compressions present in the flow. As we argue in the 
Discussion section, this is due to the higher $\mt$ in the 
former.

\section{Small-scale universal scaling}

In the classical phenomenology of turbulence,  
energy is produced at the largest scales (due to the 
geometry of the device or forcing mechanism). Non-linear
mechanisms then lead to instabilities % of these large structures 
which results in the generation of smaller and smaller scales. This 
energy transfer process continues until scales are 
small enough that the smoothing effect of 
molecular transport processes (viscous effects) become dominant and 
the energy is dissipated into heat. 
In this step-by-step energy cascade, motions progressively 
lose information from (non-universal) 
geometrical aspects at the large scales.
Thus, the hope for self-similar universality at small scales. 

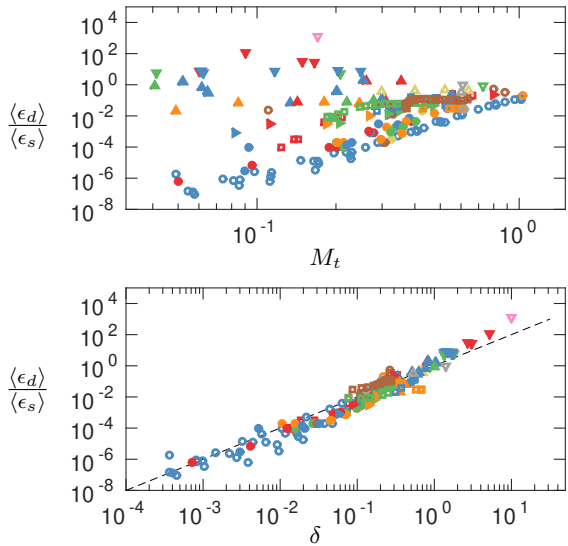
\begin{figure}
\begin{center}
\includegraphics[clip,trim=0 0 0 0,width=0.42\textwidth]{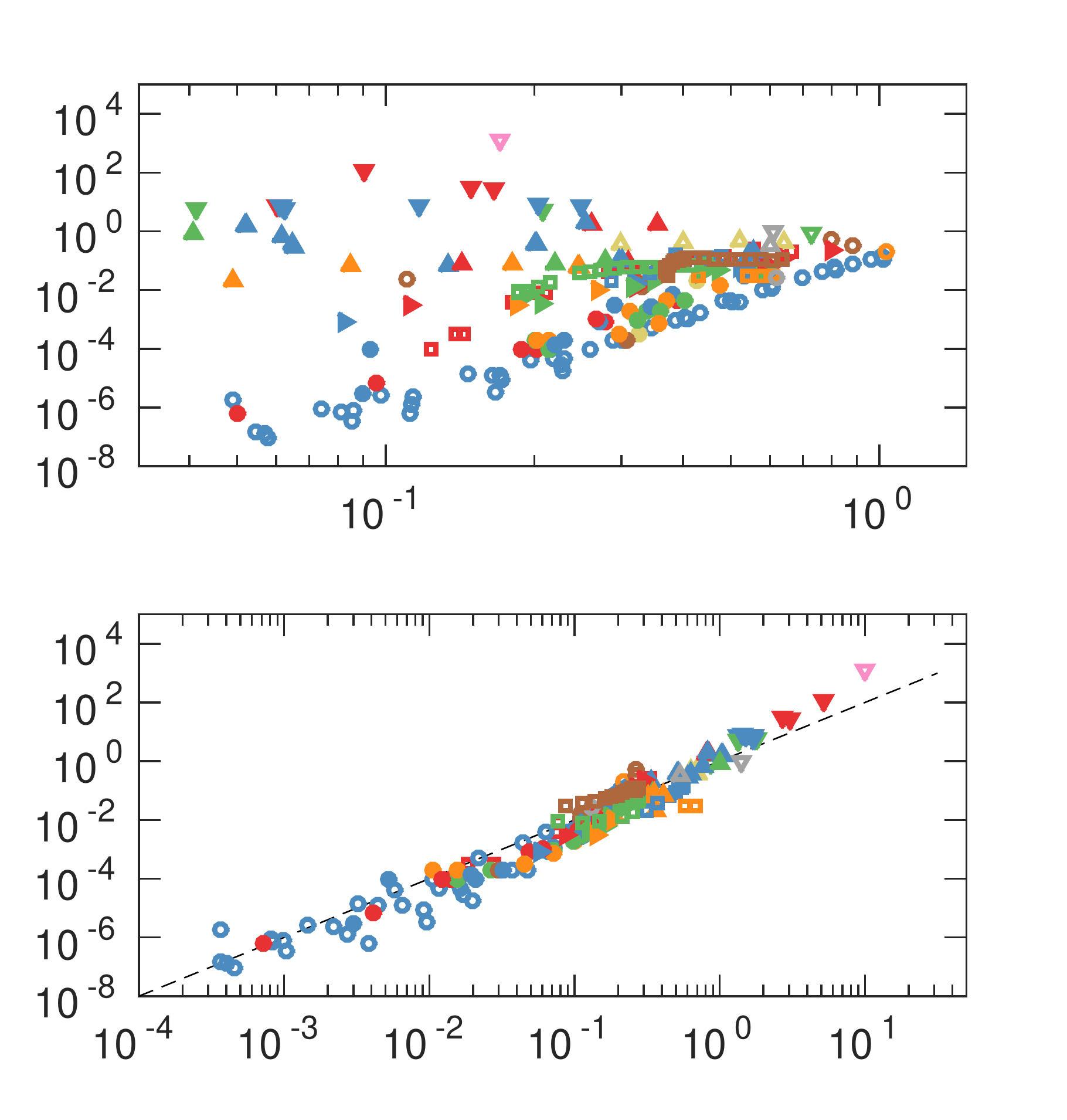}
\begin{picture}(0,0)
\put(-235,160){{\large $\frac{\avdisd}{\avdiss}$}}
\put(-235,60){{\large $\frac{\avdisd}{\avdiss}$}}
\put(-120,110){{{$M_{t}$}}}
\put(-120,5){{{$\delta$}}}
\end{picture}
\end{center}
\caption{
\label{fig:Diss}
Scaling of ratio of dilatational to solenoidal dissipation
with (a) $M_{t}$ and (b) $\delta$.
Dashed line in (b) with a slope of 2 for reference.
%A: Current simulations(HIT solenoidal forced),
%B: Current simulations(HIT dilataional forced),
%C: Wang \textit{et al.}\cite{WYSXHC2013}(HIT dilatational forced)
%D: Wang \textit{et al.}\cite{WGWPRF2017}(HIT solenoidal forced)
%E: Wang \textit{et al.}\cite{wangpf2011effect}(HIT solenoidal forced)
%F: Sarkar \textit{et al.}\cite{SEHK1991}(HST anisotropic)
%G: Kida and Orszag\cite{KO1990,KO1992} (dilataional forced isotropic)
%H: Kida and Orszag\cite{KO1990,KO1992} (solenoidal forced isotropic)
%I: Sarkar\textit{et al.}\cite{SEH1992} (HST anisotropic)
%J: Wang\textit{et al.}\cite{WJWMCSXCCS2018} (dilataional forced isotropic) 
%K: Chen \textit{et al}\cite{CWLWC2018} (HST anisotropic)
}
\end{figure}

%Compressible turbulent flows also exhibit fluctuations in 
%all hydrodynamic variables. These turbulent 
%fluctuations against the smoothing effect of molecular 
%transport processes (characterized by the macroscopic viscosity) 
%lead to dissipation mechanisms which transfer kinetic energy to
%internal energy. 
The result of this energy dissipation at small scales 
is a temperature increase and a decay 
of all spatial and temporal fluctuations. The rate at which this
happens is dictated
by the so-called energy dissipation rate
$\avdis=2\mu\la s_{ij}s_{ij}\ra$ 
(summation convention implied)
where 
$s_{ij}\equiv(\partial u_i/\partial x_j+\partial u_j/\partial x_i)/2$
is the fluctuating strain rate tensor. 
Dissipation is also 
a key ingredient in the classical understanding of turbulent flows as
it is the last step in the energy cascade from large scales to small
scales \cite{frisch95}.
In incompressible homogeneous 
flows the dissipation can also be written as 
$\avdis=\mu\la \omega_{i}\omega_{i}\ra$ 
(where $\vec{\omega}\equiv\nabla\times\vec{u}$ is the vorticity vector)
and its scaling has been the focus of a large body of literature
\cite{vassilicos2015} and is relatively well understood. 
A main result is that $\avdis$ becomes independent
of viscosity at high Reynolds numbers for the flow configurations
considered here.
In compressible flows, the situation is more complicated 
and much less is known \cite{JD2016,ED2018} due to both
the larger parameter space and the additional terms that contribute to 
dissipation. In particular, 
if the flow is homogeneous we can write $\avdis=\avdiss+\avdisd$ where
$\avdiss\equiv\la\mu \omega_i\omega_i\ra$ and 
$\avdisd\equiv(4/3)\la\mu \theta^2\ra$ with $\theta\equiv\partial u_i/\partial x_i$
being the dilatation. The first component is the so-called solenoidal dissipation 
as its expression is identical to that in incompressible flows. Clearly,
only the solenoidal component of the velocity contributes to $\avdiss$
as the dilatational component is irrotational by construction.
The second component is the so-called dilatational dissipation and is 
exclusively due to the dilatational velocity. 
The historical focus has largely been in understanding and modeling 
the dilatational dissipation as a ``correction'' 
due to compressibility \cite{GB2009}. 
The most widely used models are of the form
$\avdisd\propto \avdiss \mt^\alpha\re^\beta$ where
different models lead to different exponents 
(e.g.\ $(\alpha,\beta)=(2,0)$ \cite{SEHK1991}, $(4,-2)$ \cite{ristorcelli1997},
or $(5,0)$ \cite{WGWPRF2017})
though other more general functional forms of the type 
$\avdisd=\avdiss F(\mt)$ have been also proposed \cite{SC.book2008,GB2009}.
%\colr{[DD: John- can you check these numbers and if there are other
%models worth noting?]}.
%\textcolor{mycolwanghst2018}{[JPJ:I think this looks good 
%but was thinking if we want to add simulation data too]}. 
%\colr{[DD: You mean data showing $\mt^2$, $\mt^4$, etc? Yes]}
However, when the available data are collected together 
as in \rfig{Diss}a, it is apparent that any model following 
the similarity scaling in \req{css} will be unable 
to capture a universal behavior.
%As we suggest below, though, results can indeed be reconciled.  

Phenomenologically, since dissipation is proportional to velocity
gradients, one can estimate the scaling of each component, in view of
\req{parameters}, as 
$\avdiss\sim \mu(\usrms/L)^2g_1(\re,\mt,\delta)$
and 
$\avdisd\sim \mu(\udrms/L)^2g_2(\re,\mt,\delta)$ where 
$g_1$ and $g_2$ are some presumably universal scaling functions.
The ratio is then $\avdisd/\avdiss = \delta^2 g_3(\re,\mt,\delta)$
where $g_3$ is another universal function. In standard models, then, 
we have $g_3 \propto \re^{\alpha}\mt^{\beta}\delta^{-2}$.
To assess the scaling with the governing parameters in \rfig{Diss}b
we show the ratio of dissipation rates versus $\delta$.
We see that the all data from different flows, with different forcing schemes,
and different conditions seem to collapse along 
\be
\avdisd/\avdiss \approx \delta^2 ,
\label{eq:diss_del}
\ee
(i.e.\ $g_3\approx 1.0$).
%we plot 
%$\avdisd/\avdiss$ as a function of $\delta$.
%Thus, 
%\be
%\avdisd/\avdiss \sim \delta^2.
%\label{eq:diss_del}
%\ee
%Note that in principle the length scales for solenoidal and dilatational
%components (say $l_s$ and $l_d$) may be different,
%in which case will also have a factor 
%$(l_s/l_d)^2$ which can in turn depend on the governing parameters. 
This scaling seems robust for almost 10 order of magnitude in 
$\avdisd/\avdiss$ for all flows (including shear flows).
Whil some scatter is observed in the data, 
no distinguishable systematic trends with the other parameters
can be seen. 
This again highlights 
the inadequacy of \req{css} and adequacy of \req{css_new},
and provides strong support for broad universality
when dilatational motions are included in the set 
of governing parameters.
For later use, we note that dilatational dissipation becomes 
larger than solenoidal dissipation at $\delta\approx 0.9$.

We finally turn to the skewness of longitudinal velocity gradients
$S=\la (\partial u_1/\partial x_1)^3\ra/\la (\partial u_1/\partial x_1)^2 \ra^{3/2}$
which has also been studied extensively in incompressible flows
\cite{SA97}. 
The fact that $S$ is non-zero is a manifestation of the non-Gaussianity 
of the velocity field. Physically, 
it represent a normalized measure of the production of 
rotational motions or enstrophy ($=\la\omega^2\ra$)
due to non-linear mechanisms.
Its value is found to be an approximately 
universal constant around -0.5 for a range of incompressible flows 
at different conditions \cite{SA97}. In fact, it is so robust a
measure, that it is common to assess the emergence of ``realistic''
turbulence by examining the value of $S$ in simulations and experiments.
This practice has also been extended to compressible turbulence, though,
its interpretation is more involved since, 
at 
present, it is unknown how $S$ scale with different parameters.
Thus, values not consistent with the incompressible $S\approx -0.5$ 
%present challenges when one attempts to understand its origin:
%it 
could 
be interpreted as a compressibility effect \cite{WGWPRF2017,WSW+2011,WGWPRFshocklet2017}, 
as an indication that fully developed turbulence has 
not been achieved \cite{LLM1992},
or even as a numerical artifact \cite{WSW+2011}.
Thus, the search for universal scaling in compressible turbulence,
has also important practical implications. 

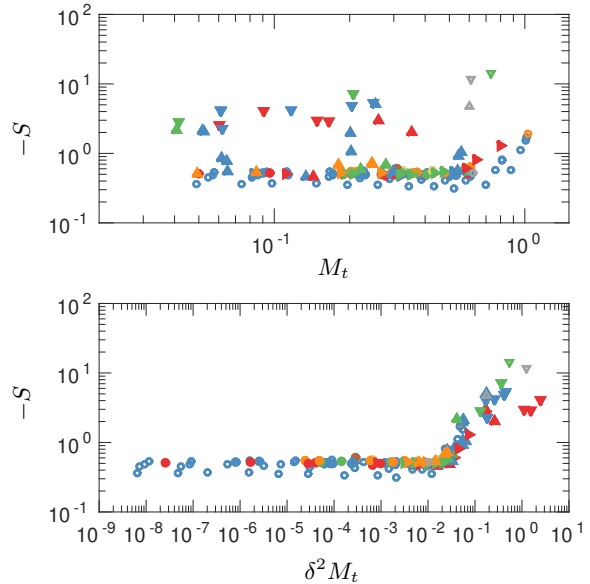
\begin{figure}
\begin{center}
\includegraphics[clip,trim=0 0 0 0,width=0.45\textwidth]{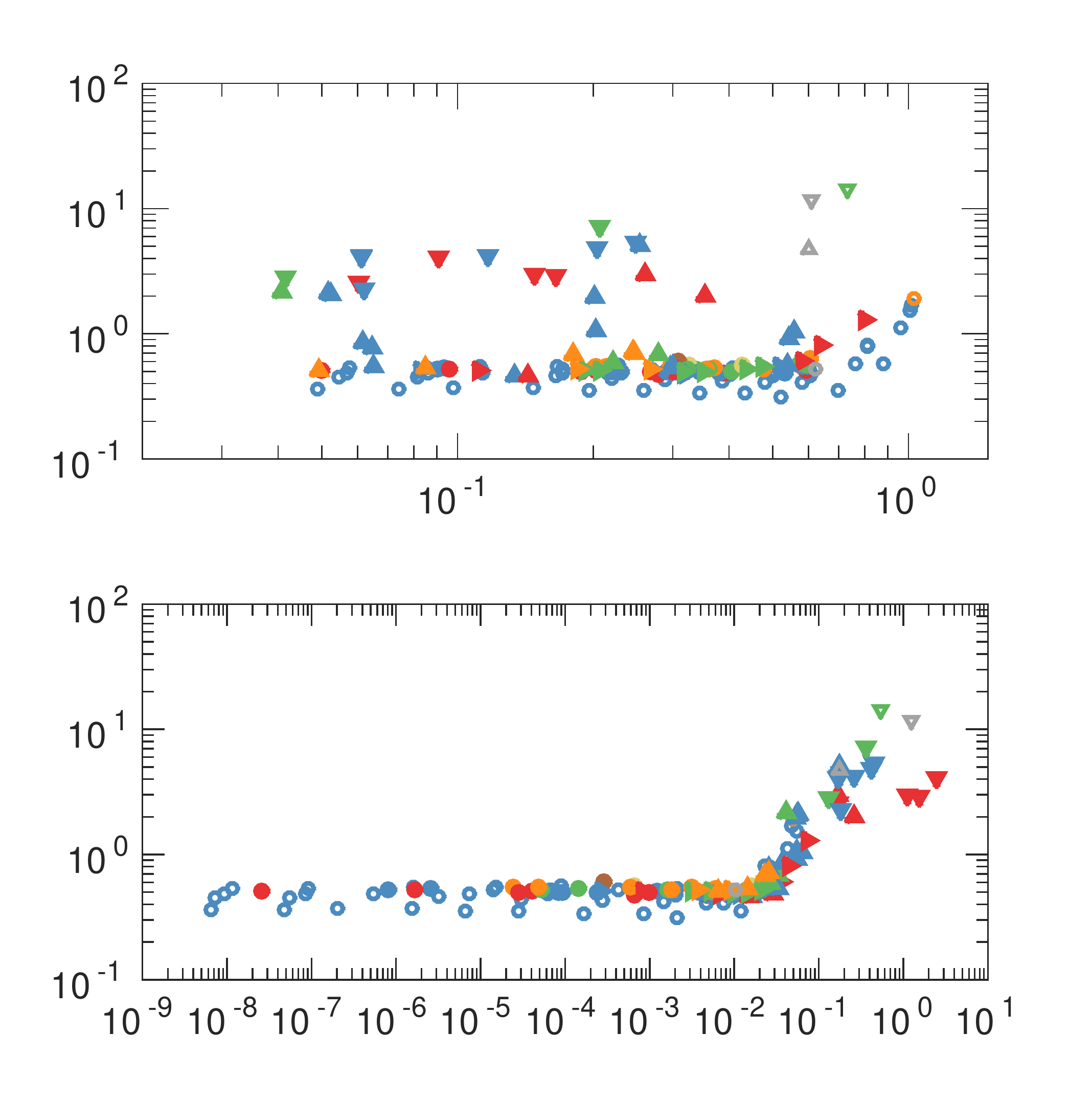}
\begin{picture}(0,0)
\put(-235,160){\rotatebox{90}{$-S$}}
\put(-120,115){$M_{t}$}
\put(-235,60){\rotatebox{90}{$-S$}}
\put(-125,0){$\delta^{2} M_{t}$}
\end{picture}
\end{center}
\caption{
\label{fig:skew}
Skewness of longitudinal velocity gradient 
% A: Current simulations(HIT solenoidal forced),
% B: Current simulations(HIT solenoidal forced),
% C: Wang \textit{et al.}\cite{WYSXHC2013}(HIT dilatational forced)
% D: Wang \textit{et al.}\cite{WGWPRF2017}(HIT solenoidal forced)
% E: Wang \textit{et al.}\cite{wangpf2011effect}(HIT solenoidal forced)
% F: Sarkar \textit{et al.}\cite{SEHK1991}(HST anisotropic)
}
\end{figure}

In \rfig{skew}a, we show the collection of $S$ as a function of 
$\mt$. 
The familiar lack of universality observed before emerges here too.
Under solenoidal forcing and low $\mt$, $S$ is consistent with 
its incompressible value which implies that 
under these conditions 
dilatational motions are either very weak 
or have equal propensity to 
form both compressions and expansions.
%\colr{[DD: Why? Can't it be 
%that its contribution is just negligible?]}.
%\textcolor{mycolwanghst2018}{[JPJ:I actually agree it can be 
%but when I wrote I was thinking about the weak equipartition regime where
%we 
%have essentially oscillations which comprises both compressions and
%expansions]}.
A number of authors \cite{WGWPRF2017,WSW+2011,WGWPRFshocklet2017}
have reported larger values of $S$ at higher $\mt$
under solenoidal forcing
which have been attributed to shocklets or small
 scale compressions \cite{lee1991}.
We see here that under dilatational forcing, 
large negative skewness is observed even at low $\mt$ which 
highlights, as before, the important
role of dilatational motions and 
the need to extend the parameter space.
Consistent with this observation is the work of 
\cite{KFKS2012} where 
it has also been observed that dilatational excitation 
leads to 
%different spatial organization of the most dissipative
%structures.
%In particular, dilatational driving leads to 
larger fractal
dimensions (corresponding to structures between
filaments and sheets) as opposed to the 
more narrow filaments observed with solenoidal forcing.

If the increased dilatational motions are responsible, through 
shocklets, of the larger value of the gradient skewness then 
we would expect wave-steepening mechanisms to be significant.
This mechanism is represented by the non-linear term 
$\ud \cdot\nabla\ud$.
To estimate the order of magnitude of this terms, 
we first note that $\ud$ is of order
$\udrms\sim \delta \us$.
The multiple scales expansion of \cite{ZM1991},
in which the solenoidal and dilatational components act at 
different scales, leads to a split gradient operator 
$\nabla=\nabla_\eta+\mt\nabla_\xi$
where $\eta$ and $\xi$ are short and long 
wavelength scales, being the latter the 
one corresponding to the acoustic contributions.
If we consider only that component 
we can estimate the order of magnitude of the 
non-linear term as $\delta^2\mt$.
%\colr{[DD: why should we take the second term?]}
%\textcolor{orange}{[JPJ:The data says us to do so in a counter intuitative 
%way that $\delta^{2} M_{t}$ corresponding to large scale scales well 
%that $\delta^{2}$ to that of small scales. As you are mentioning that 
%the Skewness divergence is due to strong gradients from other groups but what we can 
%add here is that skewness divergence is not solely a small scale phenomena but our data 
%suggests that is connected to the large scales like a cascade process. Similar to vortex stretching , 
%in incompressible turbulence is a multi scale process where what happens in large scale drives the mechanism 
%of cascade. So we can suggest that shocklets are not entirely isolated events
% at small scales enterily dissconnected from the large scales as often reported in literature 
%but might be a manifestation of the cascade process driven by large scales.
%Infact I just checked papers that suggest high skewness, none of them actually justifies 
%skewness increase with shocklets in fact they just say so ]} 
%
This phenomenological argument would imply that 
when $\delta^2\mt$ is high enough, $S$ will increase in 
magnitude.
This is indeed what we see in \rfig{skew}b where,
at low values of this parameter,
data agree with the incompressible value 
but diverges at a critical value of 
$\delta^2\mt \approx 3\times 10^{-2}$.
This $S$-divergence at a critical value of $\delta^2\mt$,
though inspired in a somewhat crude order-of-magnitude 
estimate, 
is consistent with all other available 
data in the literature.
%\cite{WGWPRF2017,WSW+2011,WGWPRFshocklet2017}.

\section{Discussion: the broader picture \label{sec:broad}}

\begin{figure}
\begin{center}
\includegraphics[clip,trim=0 0 0 0,width=0.5\textwidth]{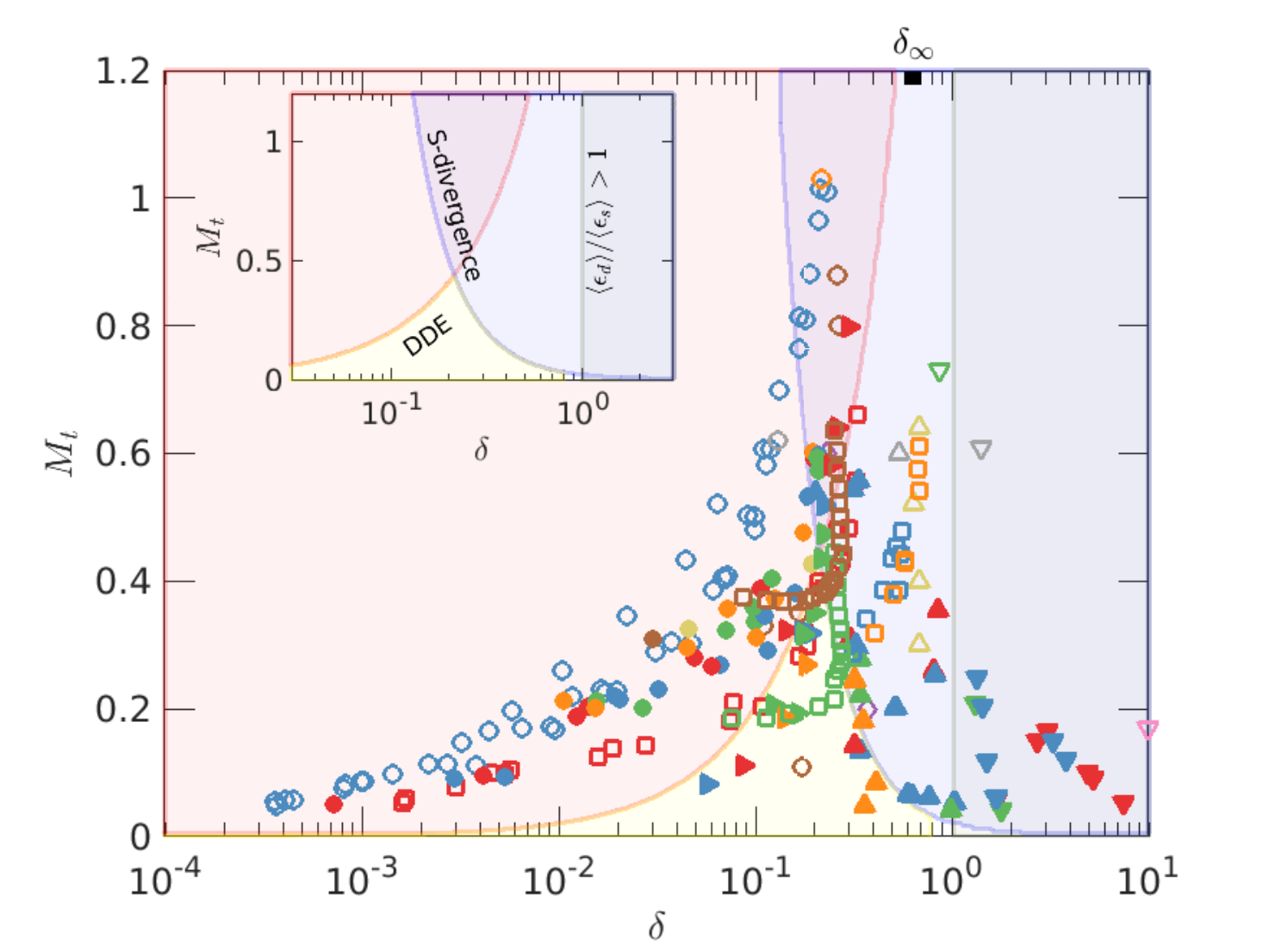}
\includegraphics[clip,trim=0 0 0 0,width=0.5\textwidth]{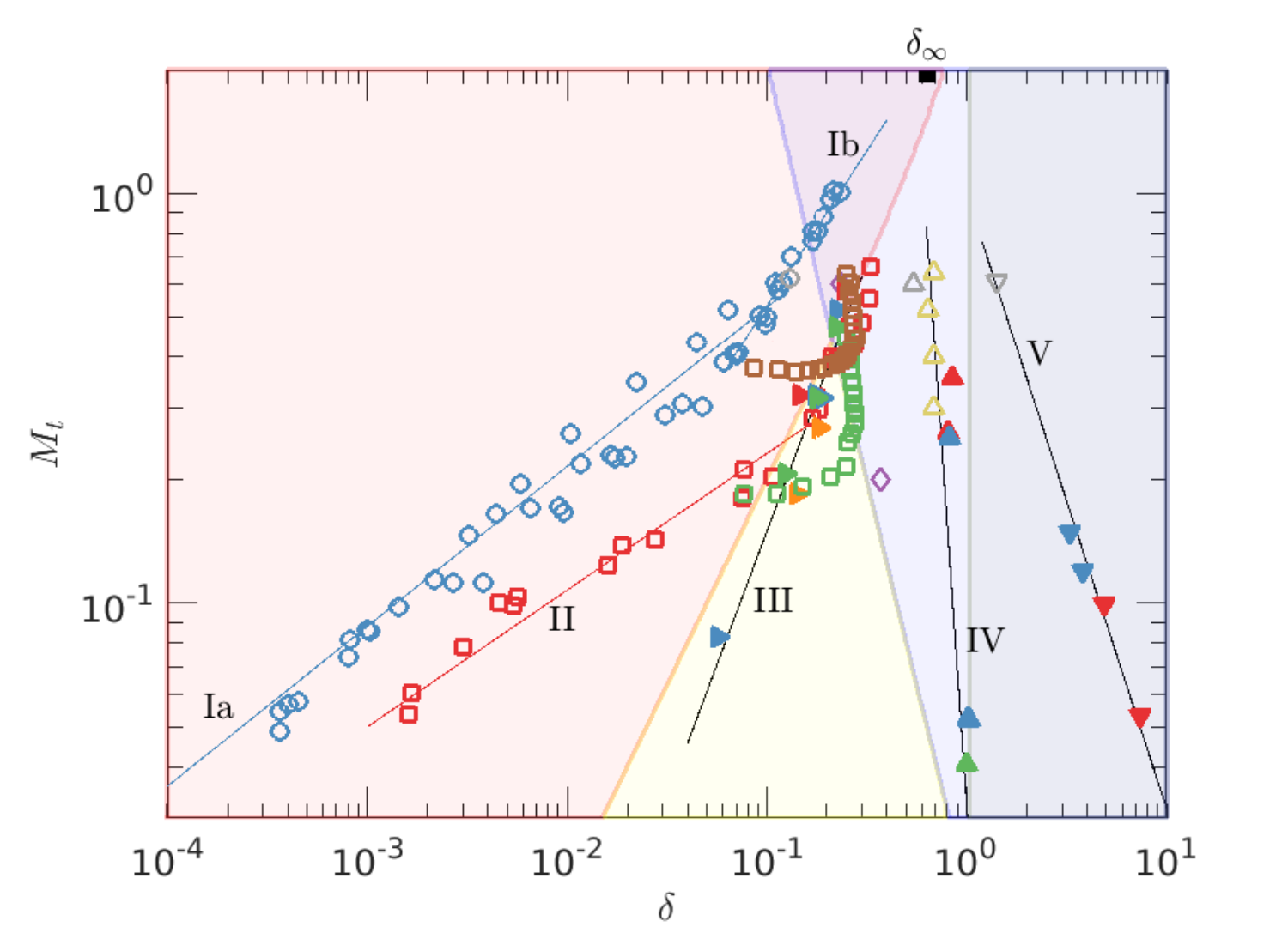}
\end{center}
\caption{
\label{fig:regions}
Regions in the $\delta$-$\mt$ diagram:
Dilatationally dominated $p$-equipartition (\DDE, ${\cal D}>{\cal D}_{crit}$),  
$S$-divergence ($\delta^2\mt> 2\times 10^{-2}$), 
and $\avdisd/\avdiss >1$.
(a) Entire database in \rtab{sources}. 
(b) Selected trajectories which include
isotropic cases with two types of solenoidal forcing, cases
with dilatational forcing, and homogeneous shear flows.
}
\end{figure}

In previous sections we showed that the statistical state 
of compressible turbulence cannot be described by the 
Reynolds and turbulent Mach numbers alone. Instead, 
a characteristic dilatational velocity needs to be incorporated 
into the governing parameters to find universal scaling laws.
This approach was shown to collapse data for fluctuations of
thermodynamic variables, dissipation rates, and skewness of 
velocity gradients.
In doing so, we identified transitions between equilibrium 
states which can be used to distinguish different regimes.
These regimes are shown in the inset of \rfig{regions}(a) with
lines separating dilatationally-dominated $p$-equipartition (\DDE),
the divergence of skewness, and the dominance of dilatational
dissipation. The main part of the figure contains all
the data from \rtab{sources} in the $\delta$-$\mt$ plane.

This figure clearly shows how $\mt$ is unable to
determine the state of turbulence in general.
Unlike classical problems in critical phenomena, 
no single transition from ``incompressible'' to ``compressible'' can be 
identified.
For example, a flow at low $\mt$
will, as $\delta$ increases, first transition to \DDE, then it will experience 
$S$-divergence and only at higher levels of $\delta$, dilatational dissipation 
will start dominating the conversion of kinetic energy to internal
energy. 
At higher $\mt$, on the other hand,
$S$-divergence may occur even before \DDE.
We see examples of both regimes in \rfig{regions}(a):
the high-$\mt$ ($\mt\gtrsim 0.8$) cases from 
\cite{WGWPRF2017} have 
much higher values of $S$ than incompressible flows
though pressure is not dominated by dilatational
dynamics while our low-$\mt$ 
($\mt\lesssim 0.4$) dilatationally forced isotropic 
cases as well as some intermediate-$\mt$ shear cases \cite{CWLWC2018},
are in \DDE\ but $S$ has not diverged yet.
Clearly, comparisons between these flows could 
lead to misleading conclusions especially if they are 
at nominally the same $\mt$.

Note that the $S$-divergence occurs, except for exceptional
conditions
(extremely low $\mt$ and high $\delta$), before dissipation is
dominated by
dilatational motions. 
If shocklets are responsible for both, as commonly 
argued, then their contribution to the third order moment of velocity
gradients emerges earlier than to the second moment. 
This observation seems to be consistent with 
recent work \cite{YD2017} suggesting that small-scale 
high-$\re$ features characterized by the anomalous scaling 
of high-order statistical moments 
in incompressible flows emerge at lower $\re$ than 
low-order moments. In subsequent work \cite{YD2018}, 
this transition from two statistical steady states 
(Gaussian at low $\re$ and anomalous at high $\re$) 
provided the key ingredients to obtain the numerical 
value of the scaling exponents 
in the anomalous regime. 
In this context, 
the discovery of a transition
such as that for $S$ (\rfig{skew}) is important 
as it identifies appropriate scaling parameters 
as well as a seemingly universal transition point. 
This could, following \cite{YD2018}, provide a fruitful 
venue to completely characterize small-scale behavior in 
compressible flows.

The collection of all the data in \rfig{regions}(a) 
spans virtually all possible regimes.
However, individually, a given flow appears to transverse 
the $\delta$-$\mt$ space along a specific trajectory.
For illustration purposes we isolate a subset of the database
in \rfig{regions}(b) comprising different flow geometries and type
of forcing: 
the present isotropic simulations with stochastic 
solenoidal and dilatational forcing,
the isotropic flows of \cite{WGWPRF2017} 
with a low-wavenumber deterministic forcing,
the homogeneous shear flow of \cite{CWLWC2018} and
\cite{sarkar1992compressible}, and isotropic simulations
with forcing that keeps the ratio of dilatation
and solenoidal energy constant at low wavenumbers
\cite{WJWMCSXCCS2018}.
%As stated in earlier sections, it is unclear
%how for a general flow described by some 
%external parameters (e.g.\ ${\cal L}$ and ${\cal U}$), 
%one can estimate solenoidal and dilatational contributions. 

From \rfig{regions}(b) we clearly see that 
different driving mechanisms or geometries generate 
different levels of dilatations.
And for a given flow, this depends on the specific 
conditions specified by the governing parameters.
%For a given flow, an increase in one 
%of the parameters, e.g.\ the strength of the random forcing,  
%could lead to a change in both $\mt$ and $\delta$ (as well as
%$\re$). In such a scenario, there may be a one-to-one 
%parametric relation between $\mt$ and $\delta$.
For example, 
as the Reynolds and Mach numbers 
are varied (at constant shear rate)
in the HST flows of \cite{CWLWC2018} (red squares),
we see an increase in both $\mt$ and $\delta$ following 
an approximate power law of the form $\mt\sim \delta^{1/3}$
(line II) or, equivalently, 
$\delta\sim \mt^3$ at low $\delta$ but appears to transition
to the \DDE\ line at ${\cal D} = 0.5$ 
(i.e.\ $\mt=0.5 \delta\sqrt{\delta^2+1}$)
as compressibility levels increase,
or to a constant $\delta$  (more below). 
This behavior is different from the 
isotropic data of \cite{WGWPRF2017} (blue circles) which 
follow $\delta\sim \mt^{2.56}$ at low $\delta$ (line Ia) 
and transitions 
to a shallower exponent $\delta\sim \mt^{1.33}$ 
(line Ib) at higher compressibility levels
never reaching \DDE. 
Compounded with this, 
the differences in the prefactors 
in the expressions relating
$\delta$ and $\mt$
result in order-of-magnitude 
differences in the value of $\delta$ for the same $\mt$.
The trajectory is also different for the isotropic 
simulation of \cite{WJWMCSXCCS2018} (yellow triangles)
which employ a forcing 
mechanism that leads to constant $\delta$ as $\mt$ is 
varied.
We also include two cases of the 
temporally evolving shear layers of 
\cite{sarkar1992compressible} (green and brown squares)
starting from different $\mt$, respectively. 
In each case the flow evolves increasing $\delta$ for $\mt$
relatively constant
at early times but approaches an approximately 
vertical line (constant $\delta$) as $\mt$ keeps increasing.
Finally we see that the dilatationally forced cases 
presented here 
(right, up and down triangles for $\sigma=10$\%, 40\% and 90\%
dilatational forcing, respectively)
present different qualitative behavior. At low dilatational
forcing ($\sigma=10\%$), $\delta$ increases as $\mt$ increases (line III)
consistent 
with the other flows, though with a smaller exponent
($\delta\sim \mt^{0.77}$).
However, for strongly dilatationally 
forced flows, the trend is the opposite with $\delta$
decreasing with $\mt$ as $\delta\sim \mt^{-0.67}$ (line V). 
At intermediate dilatational 
forcing ($\sigma=40$\%, line IV), $\delta$ is approximately constant
and close to the data of \cite{WJWMCSXCCS2018} 
which was driven by 50\% dilatational forcing.

Obviously understanding the detailed physical mechanisms 
that lead to specific trajectories in the $\delta$-$\mt$
plane is interesting in its own and their investigation 
will certainly lead to deeper 
understanding of the dynamics of particular flow 
configurations. 
Our purpose here, however,
is precisely to show that $\mt$ is insufficient to
characterize completely the statistical 
state of turbulence but that 
in combination with $\delta$ we can obtain,
for example, universal scaling laws for the 
pressure variance, energy
dissipation, and skewness of velocity gradients
regardless of the generation mechanism. 

This general view of universal statistical equilibria for compressible 
turbulence in the $\delta$-$\mt$ plane can also be used to help contextualize 
results in the literature. We have pointed out that different flows
follow different trajectories. However, it is also possible to
distinguish trajectories (or some of its features) 
that share commonalities 
for a given {\it class} of systems. For example, solenoidally
forced isotropic turbulence may follow $\delta=C \mt^\alpha$
(at least for some region in the $\delta$-$\mt$ plane)
with the same exponent $\alpha$ but different prefactors $C$
depending on the specific form of the driving mechanism.
The addition of dilatational forcing changes the
numerical value (and even sign) of the
scaling exponents $\alpha$, as discussed above.
In support of the concept of classes,
we note that data from diverse studies in the literature 
(see symbols in \rtab{sources})
using different forcing 
and numerical schemes seem to move along 
the same trajectories 
(e.g.\ lines III, IV and V).
%pointing to,
%potentially, specific classes.
Similar considerations could be applied to shear layers.
An implication of this is that, within a class, 
\req{css_new} reduces to 
\req{css} and
proposals in the literature based on $\mt$ may still 
be approximately valid. 
The investigation of these trajectories and 
identification of universal classes would require more data 
that currently available
from simulations and experiments carefully designed 
for this purpose. 
This is obviously an important 
task that would be extremely valuable in 
turbulence modeling.
The data in \rfig{regions} while limited is, in a very 
broad sense, consistent with the existence of classes.

In our search for universal features for 
compressible turbulence,
we finally consider two theoretical results.
First, Staroselsky et.\ al \cite{SYKO1990} studied isotropic 
compressible turbulence 
with a Gaussian forcing at the large scales 
using renormalization group. A
statistical equilibrium was found with a constant ratio of
solenoidal and dilatational kinetic energy
(that is, constant $\delta$) when the forcing spectrum
decays sufficiently fast with wavenumber. 
In particular, they predict  
an asymptotic limit ($\delta\to \deltai$) 
given by $\deltai^{-2}\approx 3$
or $\deltai\approx 0.58$. 
This state of equipartition, which is of different nature than 
$p$-equipartition,
will be called $\delta$-equipartition.
The second earlier study is that 
of Kraichnan \cite{kraich1955} who also suggested $\delta$-equipartition 
but based on statistical mechanics
principles. In particular, based on Liouville theorem, 
he suggested that in the inviscid case and with weak 
excitation there is equipartition between vortical 
solenoidal modes (with two degrees of freedom) 
and acoustic modes, also with two degrees of freedom
(one from the dilatational
mode and one from a bijective function of the density mode). 
Thus, $\deltai^{2} = 1/2$ or
$\deltai=0.7$, which is not far from \cite{SYKO1990}, though derived from
a completely different perspective. It is interesting that in the case
of weak fluctuations of thermodynamic variables, the more general
formulation of Kraichnan also leads to 
a form of $p$-equipartition.

The range formed by these two theoretical limits is marked on
the top of both panels of \rfig{regions} with a thick line. 
Collectively, the data seem qualitatively consistent
with such a limit as a 
universal feature across flows.
For example, we see this trend in naturally 
forced flows such as the shear
layers of \cite{sarkar1992compressible} 
where $\delta$ clearly approaches a constant 
not far from the theoretical prediction
(apparently independent of initial $\mt$)
as $\mt$ increases during the temporal evolution of the 
flow.
%as the shear layer develops.
It is also interesting that in our isotropic simulations,
when dilatational forcing is strong and kept constant, and 
temperature and viscosity are varied to achieve higher $\mt$, 
$\delta$ decreases (line IV in \rfig{regions}b).
%suggesting that at high $\mt$ high values of 
%$\delta$ may not be attainable. 
Even with purely dilatational forcing, the
values of $\delta$ seem bounded as $\mt$ increases.
Further support for such a universal asymptotic behavior 
is provided by 
the very-high-$\mt$ simulations of \cite{KGFK2012} that,
while different from all in \rtab{sources} since they are 
based on the Euler equations and the flow is isothermal, 
yield $\delta\approx 0.5$, very close to the 
the theoretical predictions, as $\mt$ is increased to values 
as high as 15. 

These observations can shed light on the proper
interpretation of the governing parameters.
In the traditional interpretation, $\mt$ is a
measure of both, separation of time scales and
compressibility level.
As we argued above this is inadequate.
Instead, $\delta$ is the appropriate non-dimensional group 
that represents compressiblity levels (dilatational motions), 
while $\mt$ represents the ratio of time scales.
When a vast disparity of time scales is present, one  
would expect little interaction between solenoidal 
and dilatational phenomena. In fact, when the equations of 
motion are decomposed into solenoidal and dilatational modes, 
the expansion of \cite{ZM1991} suggests that the cross-terms
responsible for the energy exchange between the two modes grow with $\mt$.
Thus, a simple physical picture emerges
based on a universal $\delta$-equipartition state: 
if flow conditions are such that $\delta\ne \deltai$
then an increase of $\mt$ will lead to stronger
non-linear interactions
between solenoidal and dilatational modes that would enable 
stronger redistribution of energy towards a complete
$\delta$-equipartition \cite{kraich1955}.
If $\delta<\deltai$, the stronger interaction will lead to 
transfers from the solenoidal to the dilatational mode. 
If $\delta>\deltai$ the transfer would proceed in the opposite 
direction. This phenomenological argument is 
consistent with the data
in \rfig{regions} and also with the observations
made about the stronger interaction of vortical 
motions with compression waves giving rise to 
the more curved spatial structures
in \rfig{viz} at higher $\mt$ but the same $\delta$.
Rigorous studies of energy 
transfers from the 
governing equations are certainly warranted.

It is also interesting, and perhaps counterintuitive,
to see that at high $\mt$
($\mt>\deltai\sqrt{\deltai^2+1}/{\cal D}_{crit}\approx 1.34$), 
$\deltai$ is to the left of the \DDE\ line.
Thus, if $\delta$-equipartition is indeed a 
universal asymptote at high $\mt$, then the solenoidal component 
will always dominate pressure, regardless of how
high $\mt$ is. In fact, the higher the $\mt$, the more
dominant solenoidal pressure becomes 
({$\cal D$} decreases) as one transverse 
the $\delta$-$\mt$ plane with $\delta=\deltai$.

In conclusion, using a large database of new direct numerical
simulations of isotropic compressible turbulence 
with different driving mechanisms combined with 
an extensive set of flows in the literature (including 
shear flows),
we show that universal scaling laws can 
indeed be identified for compressible turbulence if 
dilatational motions are incorporated in the 
non-dimensional groups used to determine its statistical
equilibrium state. Whereas the traditional turbulent 
Mach number fails to describe the state of the turbulence
and thus to collapse the data for different conditions and flows,
we proposed a $\delta$-$\mt$ plane in which 
different statistical equilibria can be identified,
regardless of the flow configuration and geometrical details.
Different flows transverse this plane in different ways,
but one can postulate classes of systems which 
may share universal trajectories or scaling exponents
defining those trajectories. An ultimate asymptotic regime 
predicted by renormalization group and statistical mechanics
is not inconsistent with available data.
The successful collapse of all the available the data 
(pressure variance, 
energy dissipation, enstrophy generation), 
and the identification of the proper parameters to encounter 
universal scaling laws, open the door to both 
accurate models and deeper understanding of 
compressible turbulence.

\begin{acknowledgments}
This work was partially supported by
the National Science Fundation Grant xxxxx 
and by the Air Force Office of Scientific Research Grant xxxx.
\end{acknowledgments}

\bibliography{main_aps}

%merlin.mbs apsrev4-1.bst 2010-07-25 4.21a (PWD, AO, DPC) hacked
%Control: key (0)
%Control: author (8) initials jnrlst
%Control: editor formatted (1) identically to author
%Control: production of article title (-1) disabled
%Control: page (0) single
%Control: year (1) truncated
%Control: production of eprint (0) enabled
\begin{thebibliography}{59}%
\makeatletter
\providecommand \@ifxundefined [1]{%
 \@ifx{#1\undefined}
}%
\providecommand \@ifnum [1]{%
 \ifnum #1\expandafter \@firstoftwo
 \else \expandafter \@secondoftwo
 \fi
}%
\providecommand \@ifx [1]{%
 \ifx #1\expandafter \@firstoftwo
 \else \expandafter \@secondoftwo
 \fi
}%
\providecommand \natexlab [1]{#1}%
\providecommand \enquote  [1]{``#1''}%
\providecommand \bibnamefont  [1]{#1}%
\providecommand \bibfnamefont [1]{#1}%
\providecommand \citenamefont [1]{#1}%
\providecommand \href@noop [0]{\@secondoftwo}%
\providecommand \href [0]{\begingroup \@sanitize@url \@href}%
\providecommand \@href[1]{\@@startlink{#1}\@@href}%
\providecommand \@@href[1]{\endgroup#1\@@endlink}%
\providecommand \@sanitize@url [0]{\catcode `\\12\catcode `\$12\catcode
  `\&12\catcode `\#12\catcode `\^12\catcode `\_12\catcode `\%12\relax}%
\providecommand \@@startlink[1]{}%
\providecommand \@@endlink[0]{}%
\providecommand \url  [0]{\begingroup\@sanitize@url \@url }%
\providecommand \@url [1]{\endgroup\@href {#1}{\urlprefix }}%
\providecommand \urlprefix  [0]{URL }%
\providecommand \Eprint [0]{\href }%
\providecommand \doibase [0]{http://dx.doi.org/}%
\providecommand \selectlanguage [0]{\@gobble}%
\providecommand \bibinfo  [0]{\@secondoftwo}%
\providecommand \bibfield  [0]{\@secondoftwo}%
\providecommand \translation [1]{[#1]}%
\providecommand \BibitemOpen [0]{}%
\providecommand \bibitemStop [0]{}%
\providecommand \bibitemNoStop [0]{.\EOS\space}%
\providecommand \EOS [0]{\spacefactor3000\relax}%
\providecommand \BibitemShut  [1]{\csname bibitem#1\endcsname}%
\let\auto@bib@innerbib\@empty
%</preamble>
\bibitem [{\citenamefont {Konstandin}\ \emph
  {et~al.}(2012{\natexlab{a}})\citenamefont {Konstandin}, \citenamefont
  {Girichidis}, \citenamefont {Federrath},\ and\ \citenamefont
  {Klessen}}]{KGFK2012}%
  \BibitemOpen
  \bibfield  {author} {\bibinfo {author} {\bibfnamefont {L.}~\bibnamefont
  {Konstandin}}, \bibinfo {author} {\bibfnamefont {P.}~\bibnamefont
  {Girichidis}}, \bibinfo {author} {\bibfnamefont {C.}~\bibnamefont
  {Federrath}}, \ and\ \bibinfo {author} {\bibfnamefont {R.~S.}\ \bibnamefont
  {Klessen}},\ }\href@noop {} {\bibfield  {journal} {\bibinfo  {journal} {The
  Astrophysical Journal}\ }\textbf {\bibinfo {volume} {761}},\ \bibinfo {pages}
  {149} (\bibinfo {year} {2012}{\natexlab{a}})}\BibitemShut {NoStop}%
\bibitem [{\citenamefont {Federrath}\ \emph {et~al.}(2008)\citenamefont
  {Federrath}, \citenamefont {Klessen},\ and\ \citenamefont
  {Schmidt}}]{FKS2008}%
  \BibitemOpen
  \bibfield  {author} {\bibinfo {author} {\bibfnamefont {C.}~\bibnamefont
  {Federrath}}, \bibinfo {author} {\bibfnamefont {R.~S.}\ \bibnamefont
  {Klessen}}, \ and\ \bibinfo {author} {\bibfnamefont {W.}~\bibnamefont
  {Schmidt}},\ }\href {\doibase 10.1086/595280} {\bibfield  {journal} {\bibinfo
   {journal} {Astrophys. J. Lett.}\ }\textbf {\bibinfo {volume} {688}},\
  \bibinfo {pages} {L79} (\bibinfo {year} {2008})}\BibitemShut {NoStop}%
\bibitem [{\citenamefont {Federrath}\ \emph {et~al.}(2009)\citenamefont
  {Federrath}, \citenamefont {Klessen},\ and\ \citenamefont
  {Schmidt}}]{FKS2009}%
  \BibitemOpen
  \bibfield  {author} {\bibinfo {author} {\bibfnamefont {C.}~\bibnamefont
  {Federrath}}, \bibinfo {author} {\bibfnamefont {R.~S.}\ \bibnamefont
  {Klessen}}, \ and\ \bibinfo {author} {\bibfnamefont {W.}~\bibnamefont
  {Schmidt}},\ }\href@noop {} {\bibfield  {journal} {\bibinfo  {journal}
  {Astroph. J.}\ }\textbf {\bibinfo {volume} {692}},\ \bibinfo {pages} {364}
  (\bibinfo {year} {2009})}\BibitemShut {NoStop}%
\bibitem [{\citenamefont {Wang}\ \emph {et~al.}(2018)\citenamefont {Wang},
  \citenamefont {Wan}, \citenamefont {Chen}, \citenamefont {Xie},\ and\
  \citenamefont {Chen}}]{WJWMCSXCCS2018}%
  \BibitemOpen
  \bibfield  {author} {\bibinfo {author} {\bibfnamefont {J.}~\bibnamefont
  {Wang}}, \bibinfo {author} {\bibfnamefont {M.}~\bibnamefont {Wan}}, \bibinfo
  {author} {\bibfnamefont {S.}~\bibnamefont {Chen}}, \bibinfo {author}
  {\bibfnamefont {C.}~\bibnamefont {Xie}}, \ and\ \bibinfo {author}
  {\bibfnamefont {S.}~\bibnamefont {Chen}},\ }\href {\doibase
  10.1103/PhysRevE.97.043108} {\bibfield  {journal} {\bibinfo  {journal} {Phys.
  Rev. E}\ }\textbf {\bibinfo {volume} {97}},\ \bibinfo {pages} {043108}
  (\bibinfo {year} {2018})}\BibitemShut {NoStop}%
\bibitem [{\citenamefont {Wang}\ \emph {et~al.}(2019)\citenamefont {Wang},
  \citenamefont {Wan}, \citenamefont {Chen}, \citenamefont {Xie}, \citenamefont
  {Wang},\ and\ \citenamefont {Chen}}]{CGTFJFM2019}%
  \BibitemOpen
  \bibfield  {author} {\bibinfo {author} {\bibfnamefont {J.}~\bibnamefont
  {Wang}}, \bibinfo {author} {\bibfnamefont {M.}~\bibnamefont {Wan}}, \bibinfo
  {author} {\bibfnamefont {S.}~\bibnamefont {Chen}}, \bibinfo {author}
  {\bibfnamefont {C.}~\bibnamefont {Xie}}, \bibinfo {author} {\bibfnamefont
  {L.-P.}\ \bibnamefont {Wang}}, \ and\ \bibinfo {author} {\bibfnamefont
  {S.}~\bibnamefont {Chen}},\ }\href {\doibase 10.1017/jfm.2019.116} {\bibfield
   {journal} {\bibinfo  {journal} {Journal of Fluid Mechanics}\ }\textbf
  {\bibinfo {volume} {867}},\ \bibinfo {pages} {195–215} (\bibinfo {year}
  {2019})}\BibitemShut {NoStop}%
\bibitem [{\citenamefont {Wang}\ \emph {et~al.}(2013)\citenamefont {Wang},
  \citenamefont {Yang}, \citenamefont {Shi}, \citenamefont {Xiao},
  \citenamefont {He},\ and\ \citenamefont {Chen}}]{WYSXHC2013}%
  \BibitemOpen
  \bibfield  {author} {\bibinfo {author} {\bibfnamefont {J.}~\bibnamefont
  {Wang}}, \bibinfo {author} {\bibfnamefont {Y.}~\bibnamefont {Yang}}, \bibinfo
  {author} {\bibfnamefont {Y.}~\bibnamefont {Shi}}, \bibinfo {author}
  {\bibfnamefont {Z.}~\bibnamefont {Xiao}}, \bibinfo {author} {\bibfnamefont
  {X.~T.}\ \bibnamefont {He}}, \ and\ \bibinfo {author} {\bibfnamefont
  {S.}~\bibnamefont {Chen}},\ }\href@noop {} {\bibfield  {journal} {\bibinfo
  {journal} {Journal of Turbulence}\ }\textbf {\bibinfo {volume} {14}},\
  \bibinfo {pages} {21} (\bibinfo {year} {2013})}\BibitemShut {NoStop}%
\bibitem [{\citenamefont {Wang}\ \emph
  {et~al.}(2017{\natexlab{a}})\citenamefont {Wang}, \citenamefont {Gotoh},\
  and\ \citenamefont {Watanabe}}]{WGWPRF2017}%
  \BibitemOpen
  \bibfield  {author} {\bibinfo {author} {\bibfnamefont {J.}~\bibnamefont
  {Wang}}, \bibinfo {author} {\bibfnamefont {T.}~\bibnamefont {Gotoh}}, \ and\
  \bibinfo {author} {\bibfnamefont {T.}~\bibnamefont {Watanabe}},\ }\href
  {\doibase 10.1103/PhysRevFluids.2.013403} {\bibfield  {journal} {\bibinfo
  {journal} {Phys. Rev. Fluids}\ }\textbf {\bibinfo {volume} {2}},\ \bibinfo
  {pages} {013403} (\bibinfo {year} {2017}{\natexlab{a}})}\BibitemShut
  {NoStop}%
\bibitem [{\citenamefont {Pirozzoli}\ and\ \citenamefont
  {Grasso}(2004)}]{PG2004}%
  \BibitemOpen
  \bibfield  {author} {\bibinfo {author} {\bibfnamefont {S.}~\bibnamefont
  {Pirozzoli}}\ and\ \bibinfo {author} {\bibfnamefont {F.}~\bibnamefont
  {Grasso}},\ }\href@noop {} {\bibfield  {journal} {\bibinfo  {journal} {Phys.
  Fluids}\ }\textbf {\bibinfo {volume} {16}},\ \bibinfo {pages} {4386 }
  (\bibinfo {year} {2004})}\BibitemShut {NoStop}%
\bibitem [{\citenamefont {Samtaney}\ \emph {et~al.}(2001)\citenamefont
  {Samtaney}, \citenamefont {Pullin},\ and\ \citenamefont {Kosovic}}]{SPK2001}%
  \BibitemOpen
  \bibfield  {author} {\bibinfo {author} {\bibfnamefont {R.}~\bibnamefont
  {Samtaney}}, \bibinfo {author} {\bibfnamefont {D.~I.}\ \bibnamefont
  {Pullin}}, \ and\ \bibinfo {author} {\bibfnamefont {B.}~\bibnamefont
  {Kosovic}},\ }\href@noop {} {\bibfield  {journal} {\bibinfo  {journal} {Phys.
  Fluids}\ }\textbf {\bibinfo {volume} {13}},\ \bibinfo {pages} {1415}
  (\bibinfo {year} {2001})}\BibitemShut {NoStop}%
\bibitem [{\citenamefont {Frisch}(1995)}]{frisch95}%
  \BibitemOpen
  \bibfield  {author} {\bibinfo {author} {\bibfnamefont {U.}~\bibnamefont
  {Frisch}},\ }\href@noop {} {\emph {\bibinfo {title} {Turbulence}}}\ (\bibinfo
   {publisher} {Cambridge University Press},\ \bibinfo {year}
  {1995})\BibitemShut {NoStop}%
\bibitem [{\citenamefont {Sreenivasan}(2018)}]{sreeni2018}%
  \BibitemOpen
  \bibfield  {author} {\bibinfo {author} {\bibfnamefont {K.~R.}\ \bibnamefont
  {Sreenivasan}},\ }\href@noop {} {\bibfield  {journal} {\bibinfo  {journal}
  {Proc. Nat. Acad. Sci.}\ ,\ \bibinfo {pages} {201800463}} (\bibinfo {year}
  {2018})}\BibitemShut {NoStop}%
\bibitem [{\citenamefont {Moin}\ and\ \citenamefont {Mahesh}(1998)}]{MM98}%
  \BibitemOpen
  \bibfield  {author} {\bibinfo {author} {\bibfnamefont {P.}~\bibnamefont
  {Moin}}\ and\ \bibinfo {author} {\bibfnamefont {K.}~\bibnamefont {Mahesh}},\
  }\href@noop {} {\bibfield  {journal} {\bibinfo  {journal} {Annu. Rev. Fluid
  Mech.}\ }\textbf {\bibinfo {volume} {30}},\ \bibinfo {pages} {539} (\bibinfo
  {year} {1998})}\BibitemShut {NoStop}%
\bibitem [{\citenamefont {Ishihara}\ \emph {et~al.}(2009)\citenamefont
  {Ishihara}, \citenamefont {Gotoh},\ and\ \citenamefont {Kaneda}}]{IGK2009}%
  \BibitemOpen
  \bibfield  {author} {\bibinfo {author} {\bibfnamefont {T.}~\bibnamefont
  {Ishihara}}, \bibinfo {author} {\bibfnamefont {T.}~\bibnamefont {Gotoh}}, \
  and\ \bibinfo {author} {\bibfnamefont {Y.}~\bibnamefont {Kaneda}},\
  }\href@noop {} {\bibfield  {journal} {\bibinfo  {journal} {Annu. Rev. Fluid
  Mech.}\ }\textbf {\bibinfo {volume} {41}},\ \bibinfo {pages} {165} (\bibinfo
  {year} {2009})}\BibitemShut {NoStop}%
\bibitem [{\citenamefont {Lele}(1994)}]{lele1994}%
  \BibitemOpen
  \bibfield  {author} {\bibinfo {author} {\bibfnamefont {S.~K.}\ \bibnamefont
  {Lele}},\ }\href@noop {} {\bibfield  {journal} {\bibinfo  {journal} {Annu.
  Rev. Fluid Mech.}\ }\textbf {\bibinfo {volume} {26}},\ \bibinfo {pages} {211}
  (\bibinfo {year} {1994})}\BibitemShut {NoStop}%
\bibitem [{\citenamefont {Gatski}\ and\ \citenamefont {Bonnet}(2013)}]{GB2009}%
  \BibitemOpen
  \bibfield  {author} {\bibinfo {author} {\bibfnamefont {T.~B.}\ \bibnamefont
  {Gatski}}\ and\ \bibinfo {author} {\bibfnamefont {J.-P.}\ \bibnamefont
  {Bonnet}},\ }\href@noop {} {\emph {\bibinfo {title} {Compressibility,
  turbulence and high speed flow}}},\ \bibinfo {edition} {2nd}\ ed.\ (\bibinfo
  {publisher} {Elsevier},\ \bibinfo {year} {2013})\BibitemShut {NoStop}%
\bibitem [{\citenamefont {Kida}\ and\ \citenamefont {Orszag}(1990)}]{KO1990}%
  \BibitemOpen
  \bibfield  {author} {\bibinfo {author} {\bibfnamefont {S.}~\bibnamefont
  {Kida}}\ and\ \bibinfo {author} {\bibfnamefont {S.~A.}\ \bibnamefont
  {Orszag}},\ }\href@noop {} {\bibfield  {journal} {\bibinfo  {journal} {J.
  Sci. Comp.}\ }\textbf {\bibinfo {volume} {5}},\ \bibinfo {pages} {85}
  (\bibinfo {year} {1990})}\BibitemShut {NoStop}%
\bibitem [{\citenamefont {Sarkar}(1995)}]{sarkarjfm1995}%
  \BibitemOpen
  \bibfield  {author} {\bibinfo {author} {\bibfnamefont {S.}~\bibnamefont
  {Sarkar}},\ }\href {\doibase 10.1017/S0022112095000085} {\bibfield  {journal}
  {\bibinfo  {journal} {Journal of Fluid Mechanics}\ }\textbf {\bibinfo
  {volume} {282}},\ \bibinfo {pages} {163–186} (\bibinfo {year}
  {1995})}\BibitemShut {NoStop}%
\bibitem [{\citenamefont {Vreman}\ \emph {et~al.}(1996)\citenamefont {Vreman},
  \citenamefont {Sandham},\ and\ \citenamefont {Luo}}]{vsljfm1996}%
  \BibitemOpen
  \bibfield  {author} {\bibinfo {author} {\bibfnamefont {A.~W.}\ \bibnamefont
  {Vreman}}, \bibinfo {author} {\bibfnamefont {N.~D.}\ \bibnamefont {Sandham}},
  \ and\ \bibinfo {author} {\bibfnamefont {K.~H.}\ \bibnamefont {Luo}},\ }\href
  {\doibase 10.1017/S0022112096007525} {\bibfield  {journal} {\bibinfo
  {journal} {Journal of Fluid Mechanics}\ }\textbf {\bibinfo {volume} {320}},\
  \bibinfo {pages} {235–258} (\bibinfo {year} {1996})}\BibitemShut {NoStop}%
\bibitem [{\citenamefont {Ni}(2016)}]{ni2016}%
  \BibitemOpen
  \bibfield  {author} {\bibinfo {author} {\bibfnamefont {Q.}~\bibnamefont
  {Ni}},\ }\href@noop {} {\bibfield  {journal} {\bibinfo  {journal} {Physical
  Review E}\ }\textbf {\bibinfo {volume} {93}},\ \bibinfo {pages} {043116}
  (\bibinfo {year} {2016})}\BibitemShut {NoStop}%
\bibitem [{\citenamefont {Praturi}\ and\ \citenamefont
  {Girimaji}(2019)}]{praturi2019effect}%
  \BibitemOpen
  \bibfield  {author} {\bibinfo {author} {\bibfnamefont {D.~S.}\ \bibnamefont
  {Praturi}}\ and\ \bibinfo {author} {\bibfnamefont {S.~S.}\ \bibnamefont
  {Girimaji}},\ }\href@noop {} {\bibfield  {journal} {\bibinfo  {journal}
  {Physics of Fluids}\ }\textbf {\bibinfo {volume} {31}},\ \bibinfo {pages}
  {055114} (\bibinfo {year} {2019})}\BibitemShut {NoStop}%
\bibitem [{\citenamefont {Blaisdell}\ \emph {et~al.}(1993)\citenamefont
  {Blaisdell}, \citenamefont {Mansour},\ and\ \citenamefont
  {Reynolds}}]{BMR1993}%
  \BibitemOpen
  \bibfield  {author} {\bibinfo {author} {\bibfnamefont {G.~A.}\ \bibnamefont
  {Blaisdell}}, \bibinfo {author} {\bibfnamefont {N.~N.}\ \bibnamefont
  {Mansour}}, \ and\ \bibinfo {author} {\bibfnamefont {W.~C.}\ \bibnamefont
  {Reynolds}},\ }\href {\doibase 10.1017/S0022112093002848} {\bibfield
  {journal} {\bibinfo  {journal} {J. Fluid Mech.}\ }\textbf {\bibinfo {volume}
  {256}},\ \bibinfo {pages} {443} (\bibinfo {year} {1993})}\BibitemShut
  {NoStop}%
\bibitem [{\citenamefont {Ristorcelli}\ and\ \citenamefont
  {Blaisdell}(1997)}]{RB1997}%
  \BibitemOpen
  \bibfield  {author} {\bibinfo {author} {\bibfnamefont {J.~R.}\ \bibnamefont
  {Ristorcelli}}\ and\ \bibinfo {author} {\bibfnamefont {G.~A.}\ \bibnamefont
  {Blaisdell}},\ }\href@noop {} {\bibfield  {journal} {\bibinfo  {journal}
  {Phys. Fluids}\ }\textbf {\bibinfo {volume} {9}},\ \bibinfo {pages} {4}
  (\bibinfo {year} {1997})}\BibitemShut {NoStop}%
\bibitem [{\citenamefont {Sarkar}\ \emph
  {et~al.}(1991{\natexlab{a}})\citenamefont {Sarkar}, \citenamefont
  {Erlebacher}, \citenamefont {Hussaini},\ and\ \citenamefont
  {Kreiss}}]{SEHK1991}%
  \BibitemOpen
  \bibfield  {author} {\bibinfo {author} {\bibfnamefont {S.}~\bibnamefont
  {Sarkar}}, \bibinfo {author} {\bibfnamefont {G.}~\bibnamefont {Erlebacher}},
  \bibinfo {author} {\bibfnamefont {M.~Y.}\ \bibnamefont {Hussaini}}, \ and\
  \bibinfo {author} {\bibfnamefont {H.~O.}\ \bibnamefont {Kreiss}},\
  }\href@noop {} {\bibfield  {journal} {\bibinfo  {journal} {J. Fluid Mech.}\
  }\textbf {\bibinfo {volume} {227}},\ \bibinfo {pages} {473} (\bibinfo {year}
  {1991}{\natexlab{a}})}\BibitemShut {NoStop}%
\bibitem [{\citenamefont {Kolmogorov}(1941)}]{K41}%
  \BibitemOpen
  \bibfield  {author} {\bibinfo {author} {\bibfnamefont {A.~N.}\ \bibnamefont
  {Kolmogorov}},\ }\href@noop {} {\bibfield  {journal} {\bibinfo  {journal}
  {Dokl. Akad. Nauk. SSSR}\ }\textbf {\bibinfo {volume} {30}},\ \bibinfo
  {pages} {299} (\bibinfo {year} {1941})}\BibitemShut {NoStop}%
\bibitem [{\citenamefont {Monin}\ and\ \citenamefont {Yaglom}(1975)}]{MY.II}%
  \BibitemOpen
  \bibfield  {author} {\bibinfo {author} {\bibfnamefont {A.~S.}\ \bibnamefont
  {Monin}}\ and\ \bibinfo {author} {\bibfnamefont {A.~M.}\ \bibnamefont
  {Yaglom}},\ }\href@noop {} {\emph {\bibinfo {title} {Statistical Fluid
  Mechanics, Vol. II}}}\ (\bibinfo  {publisher} {MIT Press},\ \bibinfo {year}
  {1975})\BibitemShut {NoStop}%
\bibitem [{\citenamefont {Sreenivasan}\ and\ \citenamefont
  {Antonia}(1997)}]{SA97}%
  \BibitemOpen
  \bibfield  {author} {\bibinfo {author} {\bibfnamefont {K.~R.}\ \bibnamefont
  {Sreenivasan}}\ and\ \bibinfo {author} {\bibfnamefont {R.~A.}\ \bibnamefont
  {Antonia}},\ }\href@noop {} {\bibfield  {journal} {\bibinfo  {journal} {Annu.
  Rev. Fluid Mech.}\ }\textbf {\bibinfo {volume} {29}},\ \bibinfo {pages} {435}
  (\bibinfo {year} {1997})}\BibitemShut {NoStop}%
\bibitem [{\citenamefont {Smits}\ and\ \citenamefont
  {Dussauge}(2006)}]{SD.book.2006}%
  \BibitemOpen
  \bibfield  {author} {\bibinfo {author} {\bibfnamefont {A.~J.}\ \bibnamefont
  {Smits}}\ and\ \bibinfo {author} {\bibfnamefont {J.~P.}\ \bibnamefont
  {Dussauge}},\ }\href@noop {} {\emph {\bibinfo {title} {Turbulent shear layers
  in supersonic flow}}}\ (\bibinfo  {publisher} {Springer},\ \bibinfo {year}
  {2006})\BibitemShut {NoStop}%
\bibitem [{\citenamefont {Hall}\ \emph {et~al.}(1993)\citenamefont {Hall},
  \citenamefont {Dimotakis},\ and\ \citenamefont
  {Rosemann}}]{hall1993experiments}%
  \BibitemOpen
  \bibfield  {author} {\bibinfo {author} {\bibfnamefont {J.}~\bibnamefont
  {Hall}}, \bibinfo {author} {\bibfnamefont {P.}~\bibnamefont {Dimotakis}}, \
  and\ \bibinfo {author} {\bibfnamefont {H.}~\bibnamefont {Rosemann}},\
  }\href@noop {} {\bibfield  {journal} {\bibinfo  {journal} {AIAA journal}\
  }\textbf {\bibinfo {volume} {31}},\ \bibinfo {pages} {2247} (\bibinfo {year}
  {1993})}\BibitemShut {NoStop}%
\bibitem [{\citenamefont {Dimotakis}(1991)}]{dimo1991turb}%
  \BibitemOpen
  \bibfield  {author} {\bibinfo {author} {\bibfnamefont {P.}~\bibnamefont
  {Dimotakis}},\ }\href@noop {} {\bibfield  {journal} {\bibinfo  {journal}
  {High Speed Flight Propulsion Systems}\ }\textbf {\bibinfo {volume} {137}},\
  \bibinfo {pages} {265} (\bibinfo {year} {1991})}\BibitemShut {NoStop}%
\bibitem [{\citenamefont {Freund}\ \emph {et~al.}(2000)\citenamefont {Freund},
  \citenamefont {Lele},\ and\ \citenamefont {Moin}}]{freund2000}%
  \BibitemOpen
  \bibfield  {author} {\bibinfo {author} {\bibfnamefont {J.~B.}\ \bibnamefont
  {Freund}}, \bibinfo {author} {\bibfnamefont {S.~K.}\ \bibnamefont {Lele}}, \
  and\ \bibinfo {author} {\bibfnamefont {P.}~\bibnamefont {Moin}},\ }\href@noop
  {} {\bibfield  {journal} {\bibinfo  {journal} {Journal of Fluid Mechanics}\
  }\textbf {\bibinfo {volume} {421}},\ \bibinfo {pages} {229} (\bibinfo {year}
  {2000})}\BibitemShut {NoStop}%
\bibitem [{Note1()}]{Note1}%
  \BibitemOpen
  \bibinfo {note} {A more general list of parameters would include the ratio of
  specific heats, $\gamma $ but is not included here for
  simplicity.}\BibitemShut {Stop}%
\bibitem [{\citenamefont {Jagannathan}\ and\ \citenamefont
  {Donzis}(2016)}]{JD2016}%
  \BibitemOpen
  \bibfield  {author} {\bibinfo {author} {\bibfnamefont {S.}~\bibnamefont
  {Jagannathan}}\ and\ \bibinfo {author} {\bibfnamefont {D.~A.}\ \bibnamefont
  {Donzis}},\ }\href {\doibase 10.1017/jfm.2015.754} {\bibfield  {journal}
  {\bibinfo  {journal} {J. Fluid Mech.}\ }\textbf {\bibinfo {volume} {789}},\
  \bibinfo {pages} {669} (\bibinfo {year} {2016})}\BibitemShut {NoStop}%
\bibitem [{\citenamefont {Schmidt}\ \emph {et~al.}(2009)\citenamefont
  {Schmidt}, \citenamefont {Federrath}, \citenamefont {Hupp}, \citenamefont
  {Kern},\ and\ \citenamefont {Niemeyer}}]{SFH+2009}%
  \BibitemOpen
  \bibfield  {author} {\bibinfo {author} {\bibfnamefont {W.}~\bibnamefont
  {Schmidt}}, \bibinfo {author} {\bibfnamefont {C.}~\bibnamefont {Federrath}},
  \bibinfo {author} {\bibfnamefont {M.}~\bibnamefont {Hupp}}, \bibinfo {author}
  {\bibfnamefont {S.}~\bibnamefont {Kern}}, \ and\ \bibinfo {author}
  {\bibfnamefont {J.~C.}\ \bibnamefont {Niemeyer}},\ }\href {\doibase
  10.1051/0004-6361:200809967} {\bibfield  {journal} {\bibinfo  {journal}
  {A\&A}\ }\textbf {\bibinfo {volume} {494}},\ \bibinfo {pages} {127} (\bibinfo
  {year} {2009})}\BibitemShut {NoStop}%
\bibitem [{\citenamefont {Donzis}\ and\ \citenamefont
  {Jagannathan}(2013)}]{DJ2013}%
  \BibitemOpen
  \bibfield  {author} {\bibinfo {author} {\bibfnamefont {D.~A.}\ \bibnamefont
  {Donzis}}\ and\ \bibinfo {author} {\bibfnamefont {S.}~\bibnamefont
  {Jagannathan}},\ }\href {\doibase 10.1017/jfm.2013.445} {\bibfield  {journal}
  {\bibinfo  {journal} {J. Fluid Mech.}\ }\textbf {\bibinfo {volume} {733}},\
  \bibinfo {pages} {221} (\bibinfo {year} {2013})}\BibitemShut {NoStop}%
\bibitem [{\citenamefont {Erlebacher}\ \emph {et~al.}(1990)\citenamefont
  {Erlebacher}, \citenamefont {Hussaini}, \citenamefont {Kreiss},\ and\
  \citenamefont {Sarkar}}]{EHKS1990}%
  \BibitemOpen
  \bibfield  {author} {\bibinfo {author} {\bibfnamefont {G.}~\bibnamefont
  {Erlebacher}}, \bibinfo {author} {\bibfnamefont {M.~Y.}\ \bibnamefont
  {Hussaini}}, \bibinfo {author} {\bibfnamefont {H.~O.}\ \bibnamefont
  {Kreiss}}, \ and\ \bibinfo {author} {\bibfnamefont {S.}~\bibnamefont
  {Sarkar}},\ }\href@noop {} {\bibfield  {journal} {\bibinfo  {journal}
  {Theoret. Comput. Fluid Dynamics}\ }\textbf {\bibinfo {volume} {2}},\
  \bibinfo {pages} {73} (\bibinfo {year} {1990})}\BibitemShut {NoStop}%
\bibitem [{\citenamefont {Ristorcelli}(1997)}]{ristorcelli1997}%
  \BibitemOpen
  \bibfield  {author} {\bibinfo {author} {\bibfnamefont {J.~R.}\ \bibnamefont
  {Ristorcelli}},\ }\href@noop {} {\bibfield  {journal} {\bibinfo  {journal}
  {J. Fluid Mech.}\ }\textbf {\bibinfo {volume} {347}},\ \bibinfo {pages} {37}
  (\bibinfo {year} {1997})}\BibitemShut {NoStop}%
\bibitem [{\citenamefont {Simone}\ \emph {et~al.}(1997)\citenamefont {Simone},
  \citenamefont {Coleman},\ and\ \citenamefont {Cambon}}]{SCC1997}%
  \BibitemOpen
  \bibfield  {author} {\bibinfo {author} {\bibfnamefont {A.}~\bibnamefont
  {Simone}}, \bibinfo {author} {\bibfnamefont {G.~N.}\ \bibnamefont {Coleman}},
  \ and\ \bibinfo {author} {\bibfnamefont {C.}~\bibnamefont {Cambon}},\ }\href
  {\doibase 10.1017/S0022112096003837} {\bibfield  {journal} {\bibinfo
  {journal} {J. Fluid Mech.}\ }\textbf {\bibinfo {volume} {330}},\ \bibinfo
  {pages} {307} (\bibinfo {year} {1997})}\BibitemShut {NoStop}%
\bibitem [{\citenamefont {Kida}\ and\ \citenamefont {Orszag}(1992)}]{KO1992}%
  \BibitemOpen
  \bibfield  {author} {\bibinfo {author} {\bibfnamefont {S.}~\bibnamefont
  {Kida}}\ and\ \bibinfo {author} {\bibfnamefont {S.~A.}\ \bibnamefont
  {Orszag}},\ }\href@noop {} {\bibfield  {journal} {\bibinfo  {journal} {J.
  Sci. Comp.}\ }\textbf {\bibinfo {volume} {7}},\ \bibinfo {pages} {1}
  (\bibinfo {year} {1992})}\BibitemShut {NoStop}%
\bibitem [{\citenamefont {Wang}\ \emph {et~al.}(2012)\citenamefont {Wang},
  \citenamefont {Shi}, \citenamefont {Wang}, \citenamefont {Xiao},
  \citenamefont {He},\ and\ \citenamefont {Chen}}]{WSWXHCPRL2012}%
  \BibitemOpen
  \bibfield  {author} {\bibinfo {author} {\bibfnamefont {J.}~\bibnamefont
  {Wang}}, \bibinfo {author} {\bibfnamefont {Y.}~\bibnamefont {Shi}}, \bibinfo
  {author} {\bibfnamefont {L.-P.}\ \bibnamefont {Wang}}, \bibinfo {author}
  {\bibfnamefont {Z.}~\bibnamefont {Xiao}}, \bibinfo {author} {\bibfnamefont
  {X.~T.}\ \bibnamefont {He}}, \ and\ \bibinfo {author} {\bibfnamefont
  {S.}~\bibnamefont {Chen}},\ }\href {\doibase 10.1103/PhysRevLett.108.214505}
  {\bibfield  {journal} {\bibinfo  {journal} {Phys. Rev. Lett.}\ }\textbf
  {\bibinfo {volume} {108}},\ \bibinfo {pages} {214505} (\bibinfo {year}
  {2012})}\BibitemShut {NoStop}%
\bibitem [{\citenamefont {Wang}\ \emph {et~al.}(2011)\citenamefont {Wang},
  \citenamefont {Shi}, \citenamefont {Wang}, \citenamefont {Xiao},
  \citenamefont {He},\ and\ \citenamefont {Chen}}]{WSW+2011}%
  \BibitemOpen
  \bibfield  {author} {\bibinfo {author} {\bibfnamefont {J.}~\bibnamefont
  {Wang}}, \bibinfo {author} {\bibfnamefont {Y.}~\bibnamefont {Shi}}, \bibinfo
  {author} {\bibfnamefont {L.-P.}\ \bibnamefont {Wang}}, \bibinfo {author}
  {\bibfnamefont {Z.}~\bibnamefont {Xiao}}, \bibinfo {author} {\bibfnamefont
  {X.}~\bibnamefont {He}}, \ and\ \bibinfo {author} {\bibfnamefont
  {S.}~\bibnamefont {Chen}},\ }\href@noop {} {\bibfield  {journal} {\bibinfo
  {journal} {Phys. Fluids}\ }\textbf {\bibinfo {volume} {23}},\ \bibinfo
  {pages} {125103} (\bibinfo {year} {2011})}\BibitemShut {NoStop}%
\bibitem [{\citenamefont {Chen}\ \emph {et~al.}(2018)\citenamefont {Chen},
  \citenamefont {Wang}, \citenamefont {Li}, \citenamefont {Wan},\ and\
  \citenamefont {Chen}}]{CWLWC2018}%
  \BibitemOpen
  \bibfield  {author} {\bibinfo {author} {\bibfnamefont {S.}~\bibnamefont
  {Chen}}, \bibinfo {author} {\bibfnamefont {J.}~\bibnamefont {Wang}}, \bibinfo
  {author} {\bibfnamefont {H.}~\bibnamefont {Li}}, \bibinfo {author}
  {\bibfnamefont {M.}~\bibnamefont {Wan}}, \ and\ \bibinfo {author}
  {\bibfnamefont {S.}~\bibnamefont {Chen}},\ }\href@noop {} {\bibfield
  {journal} {\bibinfo  {journal} {Physics of Fluids}\ }\textbf {\bibinfo
  {volume} {30}},\ \bibinfo {pages} {065109} (\bibinfo {year}
  {2018})}\BibitemShut {NoStop}%
\bibitem [{\citenamefont {Sarkar}\ \emph
  {et~al.}(1991{\natexlab{b}})\citenamefont {Sarkar}, \citenamefont
  {Erlebacher},\ and\ \citenamefont {Hussaini}}]{SEHYTCFD1991}%
  \BibitemOpen
  \bibfield  {author} {\bibinfo {author} {\bibfnamefont {S.}~\bibnamefont
  {Sarkar}}, \bibinfo {author} {\bibfnamefont {G.}~\bibnamefont {Erlebacher}},
  \ and\ \bibinfo {author} {\bibfnamefont {M.~Y.}\ \bibnamefont {Hussaini}},\
  }\href@noop {} {\bibfield  {journal} {\bibinfo  {journal} {Theoretical and
  Computational Fluid Dynamics}\ }\textbf {\bibinfo {volume} {2}},\ \bibinfo
  {pages} {291} (\bibinfo {year} {1991}{\natexlab{b}})}\BibitemShut {NoStop}%
\bibitem [{\citenamefont {Sarkar}\ \emph {et~al.}(1993)\citenamefont {Sarkar},
  \citenamefont {Erlebacher},\ and\ \citenamefont
  {Hussaini}}]{sarkar1992compressible}%
  \BibitemOpen
  \bibfield  {author} {\bibinfo {author} {\bibfnamefont {S.}~\bibnamefont
  {Sarkar}}, \bibinfo {author} {\bibfnamefont {G.}~\bibnamefont {Erlebacher}},
  \ and\ \bibinfo {author} {\bibfnamefont {M.}~\bibnamefont {Hussaini}},\
  }\enquote {\bibinfo {title} {Compressible homogeneous shear: simulation and
  modeling. in turbulent shear flows 8 (ed. f.durst et al.).}}\ in\ \href@noop
  {} {\emph {\bibinfo {booktitle} {Turbulent Shear Flows 8}}}\ (\bibinfo
  {publisher} {Springer},\ \bibinfo {year} {1993})\ pp.\ \bibinfo {pages}
  {249--267}\BibitemShut {NoStop}%
\bibitem [{\citenamefont {Sagaut}\ and\ \citenamefont
  {Cambon}(2008)}]{SC.book2008}%
  \BibitemOpen
  \bibfield  {author} {\bibinfo {author} {\bibfnamefont {P.}~\bibnamefont
  {Sagaut}}\ and\ \bibinfo {author} {\bibfnamefont {C.}~\bibnamefont
  {Cambon}},\ }\href@noop {} {\emph {\bibinfo {title} {Homogeneous Turbulence
  Dynamics}}}\ (\bibinfo  {publisher} {Cambridge University Press},\ \bibinfo
  {address} {Cambridge},\ \bibinfo {year} {2008})\BibitemShut {NoStop}%
\bibitem [{\citenamefont {Landau}\ and\ \citenamefont
  {Lifshitz}(1987)}]{LL.book.1987}%
  \BibitemOpen
  \bibfield  {author} {\bibinfo {author} {\bibfnamefont {L.~D.}\ \bibnamefont
  {Landau}}\ and\ \bibinfo {author} {\bibfnamefont {E.~M.}\ \bibnamefont
  {Lifshitz}},\ }\href@noop {} {\emph {\bibinfo {title} {Fluid Mechanics}}}\
  (\bibinfo  {publisher} {Pergamon Press},\ \bibinfo {year} {1987})\BibitemShut
  {NoStop}%
\bibitem [{\citenamefont {Donzis}\ \emph {et~al.}(2012)\citenamefont {Donzis},
  \citenamefont {Sreenivasan},\ and\ \citenamefont {Yeung}}]{DSY2012}%
  \BibitemOpen
  \bibfield  {author} {\bibinfo {author} {\bibfnamefont {D.~A.}\ \bibnamefont
  {Donzis}}, \bibinfo {author} {\bibfnamefont {K.~R.}\ \bibnamefont
  {Sreenivasan}}, \ and\ \bibinfo {author} {\bibfnamefont {P.~K.}\ \bibnamefont
  {Yeung}},\ }\href@noop {} {\bibfield  {journal} {\bibinfo  {journal} {Physica
  D}\ }\textbf {\bibinfo {volume} {241}},\ \bibinfo {pages} {164 } (\bibinfo
  {year} {2012})}\BibitemShut {NoStop}%
\bibitem [{\citenamefont {Barenblatt}\ and\ \citenamefont
  {Zeldovich}(1972)}]{BZ1972}%
  \BibitemOpen
  \bibfield  {author} {\bibinfo {author} {\bibfnamefont {G.~I.}\ \bibnamefont
  {Barenblatt}}\ and\ \bibinfo {author} {\bibfnamefont {Y.~B.}\ \bibnamefont
  {Zeldovich}},\ }\href@noop {} {\bibfield  {journal} {\bibinfo  {journal}
  {Annu. Rev. Fluid Mech.}\ }\textbf {\bibinfo {volume} {4}},\ \bibinfo {pages}
  {285} (\bibinfo {year} {1972})}\BibitemShut {NoStop}%
\bibitem [{\citenamefont {Wang}\ \emph
  {et~al.}(2017{\natexlab{b}})\citenamefont {Wang}, \citenamefont {Gotoh},\
  and\ \citenamefont {Watanabe}}]{WGWPF2017inter}%
  \BibitemOpen
  \bibfield  {author} {\bibinfo {author} {\bibfnamefont {J.}~\bibnamefont
  {Wang}}, \bibinfo {author} {\bibfnamefont {T.}~\bibnamefont {Gotoh}}, \ and\
  \bibinfo {author} {\bibfnamefont {T.}~\bibnamefont {Watanabe}},\ }\href
  {\doibase 10.1103/PhysRevFluids.2.053401} {\bibfield  {journal} {\bibinfo
  {journal} {Phys. Rev. Fluids}\ }\textbf {\bibinfo {volume} {2}},\ \bibinfo
  {pages} {053401} (\bibinfo {year} {2017}{\natexlab{b}})}\BibitemShut
  {NoStop}%
\bibitem [{\citenamefont {Vassilicos}(2015)}]{vassilicos2015}%
  \BibitemOpen
  \bibfield  {author} {\bibinfo {author} {\bibfnamefont {J.~C.}\ \bibnamefont
  {Vassilicos}},\ }\href@noop {} {\bibfield  {journal} {\bibinfo  {journal}
  {Annual Review of Fluid Mechanics}\ }\textbf {\bibinfo {volume} {47}},\
  \bibinfo {pages} {95} (\bibinfo {year} {2015})}\BibitemShut {NoStop}%
\bibitem [{\citenamefont {Eyink}\ and\ \citenamefont {Drivas}(2018)}]{ED2018}%
  \BibitemOpen
  \bibfield  {author} {\bibinfo {author} {\bibfnamefont {G.~L.}\ \bibnamefont
  {Eyink}}\ and\ \bibinfo {author} {\bibfnamefont {T.~D.}\ \bibnamefont
  {Drivas}},\ }\href {\doibase 10.1103/PhysRevX.8.011022} {\bibfield  {journal}
  {\bibinfo  {journal} {Phys. Rev. X}\ }\textbf {\bibinfo {volume} {8}},\
  \bibinfo {pages} {011022} (\bibinfo {year} {2018})}\BibitemShut {NoStop}%
\bibitem [{\citenamefont {Wang}\ \emph
  {et~al.}(2017{\natexlab{c}})\citenamefont {Wang}, \citenamefont {Gotoh},\
  and\ \citenamefont {Watanabe}}]{WGWPRFshocklet2017}%
  \BibitemOpen
  \bibfield  {author} {\bibinfo {author} {\bibfnamefont {J.}~\bibnamefont
  {Wang}}, \bibinfo {author} {\bibfnamefont {T.}~\bibnamefont {Gotoh}}, \ and\
  \bibinfo {author} {\bibfnamefont {T.}~\bibnamefont {Watanabe}},\ }\href
  {\doibase 10.1103/PhysRevFluids.2.023401} {\bibfield  {journal} {\bibinfo
  {journal} {Phys. Rev. Fluids}\ }\textbf {\bibinfo {volume} {2}},\ \bibinfo
  {pages} {023401} (\bibinfo {year} {2017}{\natexlab{c}})}\BibitemShut
  {NoStop}%
\bibitem [{\citenamefont {Lee}\ \emph {et~al.}(1992)\citenamefont {Lee},
  \citenamefont {Lele},\ and\ \citenamefont {Moin}}]{LLM1992}%
  \BibitemOpen
  \bibfield  {author} {\bibinfo {author} {\bibfnamefont {S.}~\bibnamefont
  {Lee}}, \bibinfo {author} {\bibfnamefont {S.~K.}\ \bibnamefont {Lele}}, \
  and\ \bibinfo {author} {\bibfnamefont {P.}~\bibnamefont {Moin}},\ }\href@noop
  {} {\bibfield  {journal} {\bibinfo  {journal} {Phys. Fluids}\ }\textbf
  {\bibinfo {volume} {4}},\ \bibinfo {pages} {1521} (\bibinfo {year}
  {1992})}\BibitemShut {NoStop}%
\bibitem [{\citenamefont {Lee}\ \emph {et~al.}(1991)\citenamefont {Lee},
  \citenamefont {Lele},\ and\ \citenamefont {Moin}}]{lee1991}%
  \BibitemOpen
  \bibfield  {author} {\bibinfo {author} {\bibfnamefont {S.}~\bibnamefont
  {Lee}}, \bibinfo {author} {\bibfnamefont {S.~K.}\ \bibnamefont {Lele}}, \
  and\ \bibinfo {author} {\bibfnamefont {P.}~\bibnamefont {Moin}},\ }\href@noop
  {} {\bibfield  {journal} {\bibinfo  {journal} {Phys. Fluids}\ }\textbf
  {\bibinfo {volume} {3}},\ \bibinfo {pages} {657} (\bibinfo {year}
  {1991})}\BibitemShut {NoStop}%
\bibitem [{\citenamefont {Konstandin}\ \emph
  {et~al.}(2012{\natexlab{b}})\citenamefont {Konstandin}, \citenamefont
  {Federrath}, \citenamefont {Klessen},\ and\ \citenamefont
  {Schmidt}}]{KFKS2012}%
  \BibitemOpen
  \bibfield  {author} {\bibinfo {author} {\bibfnamefont {L.}~\bibnamefont
  {Konstandin}}, \bibinfo {author} {\bibfnamefont {C.}~\bibnamefont
  {Federrath}}, \bibinfo {author} {\bibfnamefont {R.~S.}\ \bibnamefont
  {Klessen}}, \ and\ \bibinfo {author} {\bibfnamefont {W.}~\bibnamefont
  {Schmidt}},\ }\href@noop {} {\bibfield  {journal} {\bibinfo  {journal} {J.
  Fluid Mech.}\ }\textbf {\bibinfo {volume} {692}},\ \bibinfo {pages} {183}
  (\bibinfo {year} {2012}{\natexlab{b}})}\BibitemShut {NoStop}%
\bibitem [{\citenamefont {Zank}\ and\ \citenamefont
  {Matthaeus}(1991)}]{ZM1991}%
  \BibitemOpen
  \bibfield  {author} {\bibinfo {author} {\bibfnamefont {G.~P.}\ \bibnamefont
  {Zank}}\ and\ \bibinfo {author} {\bibfnamefont {W.~H.}\ \bibnamefont
  {Matthaeus}},\ }\href@noop {} {\bibfield  {journal} {\bibinfo  {journal}
  {Phys. Fluids}\ }\textbf {\bibinfo {volume} {3}},\ \bibinfo {pages} {69}
  (\bibinfo {year} {1991})}\BibitemShut {NoStop}%
\bibitem [{\citenamefont {Yakhot}\ and\ \citenamefont {Donzis}(2017)}]{YD2017}%
  \BibitemOpen
  \bibfield  {author} {\bibinfo {author} {\bibfnamefont {V.}~\bibnamefont
  {Yakhot}}\ and\ \bibinfo {author} {\bibfnamefont {D.~A.}\ \bibnamefont
  {Donzis}},\ }\href {\doibase 10.1103/PhysRevLett.119.044501} {\bibfield
  {journal} {\bibinfo  {journal} {Phys. Rev. Lett.}\ }\textbf {\bibinfo
  {volume} {119}},\ \bibinfo {pages} {044501} (\bibinfo {year}
  {2017})}\BibitemShut {NoStop}%
\bibitem [{\citenamefont {Yakhot}\ and\ \citenamefont {Donzis}(2018)}]{YD2018}%
  \BibitemOpen
  \bibfield  {author} {\bibinfo {author} {\bibfnamefont {V.}~\bibnamefont
  {Yakhot}}\ and\ \bibinfo {author} {\bibfnamefont {D.~A.}\ \bibnamefont
  {Donzis}},\ }\href {\doibase https://doi.org/10.1016/j.physd.2018.07.005}
  {\bibfield  {journal} {\bibinfo  {journal} {Physica D.}\ }\textbf {\bibinfo
  {volume} {384-385}},\ \bibinfo {pages} {12} (\bibinfo {year}
  {2018})}\BibitemShut {NoStop}%
\bibitem [{\citenamefont {Staroselsky}\ \emph {et~al.}(1990)\citenamefont
  {Staroselsky}, \citenamefont {Yakhot}, \citenamefont {Kida},\ and\
  \citenamefont {Orszag}}]{SYKO1990}%
  \BibitemOpen
  \bibfield  {author} {\bibinfo {author} {\bibfnamefont {I.}~\bibnamefont
  {Staroselsky}}, \bibinfo {author} {\bibfnamefont {V.}~\bibnamefont {Yakhot}},
  \bibinfo {author} {\bibfnamefont {S.}~\bibnamefont {Kida}}, \ and\ \bibinfo
  {author} {\bibfnamefont {S.~A.}\ \bibnamefont {Orszag}},\ }\href {\doibase
  10.1103/PhysRevLett.65.171} {\bibfield  {journal} {\bibinfo  {journal} {Phys.
  Rev. Lett.}\ }\textbf {\bibinfo {volume} {65}},\ \bibinfo {pages} {171}
  (\bibinfo {year} {1990})}\BibitemShut {NoStop}%
\bibitem [{\citenamefont {Kraichnan}(1955)}]{kraich1955}%
  \BibitemOpen
  \bibfield  {author} {\bibinfo {author} {\bibfnamefont {R.~H.}\ \bibnamefont
  {Kraichnan}},\ }\href {\doibase http://dx.doi.org/10.1121/1.1907924}
  {\bibfield  {journal} {\bibinfo  {journal} {J. Acous. Soc. Amer.}\ }\textbf
  {\bibinfo {volume} {27}},\ \bibinfo {pages} {438} (\bibinfo {year}
  {1955})}\BibitemShut {NoStop}%
\end{thebibliography}%

%\IfFileExists{pnas-new.bst}{
%  \bibliographystyle{pnas-new}
%}

%\begin{thebibliography}{10}
%\bibitem{BN}
%M.~Belkin and P.~Niyogi, {\em Using manifold structure for partially
%  labelled classification}, Advances in NIPS, 15 (2003).
%
%\bibitem{BBG:EmbeddingRiemannianManifoldHeatKernel}
%P.~B\'erard, G.~Besson, and S.~Gallot, {\em Embedding {R}iemannian
%  manifolds by their heat kernel}, Geom. and Fun. Anal., 4 (1994),
%  pp.~374--398.
%
%\end{thebibliography}

%\end{article}

\end{document}